\documentclass[12pt]{article}
\usepackage[margin=1.0in]{geometry}

\DeclareUnicodeCharacter{2212}{-} \DeclareUnicodeCharacter{FB01}{-} \DeclareUnicodeCharacter{FB02}{-}

\usepackage[normalem]{ulem} 

\usepackage{amsmath}
\usepackage{amsfonts}
\usepackage{amssymb}
\usepackage{bm, bbm} 
\usepackage{bbold}
\usepackage{subcaption}
\usepackage{graphicx}
\usepackage{xcolor}
\usepackage{hyperref}
\usepackage{cleveref}
\usepackage{cite}
\usepackage{comment}
\usepackage{multirow}
\usepackage{tikz}
\usetikzlibrary{shapes.geometric, arrows.meta, positioning,decorations.pathreplacing}
\usepackage[T1]{fontenc}
\usepackage[utf8]{inputenc}
\usepackage[greek,english]{babel}

\newcommand{\be}{\begin{equation}}
\newcommand{\ee}{\end{equation}}

\usepackage{xcolor}
\usepackage{bm}
\usepackage{titling}
\usepackage{tikz}
\usetikzlibrary{calc}

\def\bea{\begin{eqnarray}}
\def\eea{\end{eqnarray}}
\def\l{\left}
\def\r{\right}

\def\nn{\nonumber\\}

\title{Towards a Post-Inflationary Composite Axion Model}

\author{Aleksandr Azatov,${}^{1,2,3}$ Mohamed Mahdi Khalil,${}^{1,2}$ and Motoo Suzuki${}^{1,2,3}$ \\[10pt]
{\small\it ${}^1$ SISSA, Via Bonomea 265, Trieste 34136, Italy}\\
{\small\it ${}^2$INFN - Sezione di Trieste, Via Valerio 2, 34127, Italy}\\
{\small\it ${}^3$IFPU, Via Beirut 2, 34014 Trieste, Italy}\\
}

\pretitle{%
  \begin{flushright}\small
 SISSA 13/2025/FISI\\
  \end{flushright}
  \vspace{1.5em}%
  \begin{center}\LARGE%
}
\posttitle{\end{center}\vspace{0.5em}}

\begin{document}

\maketitle


\begin{abstract}
Composite axions offer a scenario where the axion emerges as a pion-like state, avoiding fine-tuning of elementary scalars and ameliorating the axion quality problem. Despite these advantages, their post-inflationary cosmology remains largely unexplored, with challenges including the domain wall problem and the presence of exotic relics. We propose two composite axion models with an effective domain wall number $N_\text{DW} = 1$ and study the dilution of relics via a short period of inflation.  
One model is based on an $SU(5)$ chiral gauge theory, while the other employs a ``gauged’’ $U(1)$ Peccei-Quinn symmetry in vector-like $SU(N)$ gauge theories.
We identify the viable parameter space in which axion strings re-enter the horizon  before or even after the QCD transition and axion dark matter is dominantly produced from the decay of the string–wall network.
\end{abstract}

\newpage
\tableofcontents
\newpage
\section{Introduction}
Despite its remarkable success, the Standard Model of particle physics leaves several fundamental questions unresolved. 
Among them, the strong CP problem stands out as a particularly puzzling fine-tuning issue. The unnaturally small value of the QCD $\theta$-angle, bounded by experimental limits to be less than $10^{-10}$~\cite{Baker:2006ts,Pendlebury:2015lrz,Abel:2020pzs}, is hard to accept without invoking a deeper structural reason.
One of the most compelling solutions is the Peccei-Quinn mechanism~\cite{Peccei:1977hh}, which introduces an axion field~\cite{Weinberg:1977ma,Wilczek:1977pj} whose dynamics relaxes the $\theta$-parameter to zero. Remarkably, the axion can simultaneously account for the observed dark matter abundance~\cite{Preskill:1982cy,Abbott:1982af,Dine:1982ah}. Yet, the story does not end there: even axion models face further challenges, most notably the axion quality problem~\cite{Georgi:1981pu,Kamionkowski:1992mf,Holman:1992us,Kallosh:1995hi,Barr:1992qq,Ghigna:1992iv,Dine:1992vx}, which questions the robustness of the Peccei-Quinn symmetry against Planck-suppressed operators.

Among various extensions of axion models, the composite axion offers a particularly intriguing possibility, in which the axion emerges as a pion-like bound state from confining strong dynamics.
In addition to solving the strong CP problem, composite axion models naturally accommodate an axion decay constant well below the Planck scale, thereby avoiding the fine-tuning issues typically associated with elementary scalar fields. Such a lower decay constant is also cosmologically motivated, as it can yield the correct axion dark matter abundance.
Early realizations along these lines were proposed by Kim and Choi~\cite{Kim:1984pt,Choi:1985cb}, based on vector-like gauge theories. However, these models generally suffer from the axion quality problem, since gauge-invariant mass terms are allowed unless one imposes ad hoc global Peccei–Quinn symmetries.
More recently, attention has shifted to models based on chiral gauge theories such as~\cite{Randall:1992ut,Redi:2016esr,Lillard:2017cwx,Lee:2018yak,Gavela:2018paw,Vecchi:2021shj,Contino:2021ayn,Cox:2023dou,Gherghetta:2025fip,Gherghetta:2025kff,Sato:2025rok,Agrawal:2025mke}, where the Peccei–Quinn symmetry can emerge as an accidental symmetry of the gauge dynamics. This approach offers a promising path to addressing the quality problem without introducing arbitrary global symmetries.

In spite of the development of the composite axion models, their cosmological implications--especially in the post-inflationary scenario--remains largely unexplored. This is primarily due to two major obstacles~\cite{Lu:2023ayc}. First, many composite axion constructions predict a domain wall number $N_{\rm DW}>1$, leading to the formation of stable domain walls that conflict with standard cosmology.
Second, the strong dynamics often give rise to heavy states charged under QCD, some of which remain fractionally charged under electromagnetism at low energies. The cosmological abundance of such exotic relics is tightly constrained by observations, posing a serious challenge for post-inflationary composite axion scenarios.
Moreover, attempts to solve the quality problem can make the relic issue more severe by enhancing the stability of unwanted heavy states.

In this paper, we take a step toward realizing a viable post-inflationary 
cosmological scenario by constructing two explicit examples of composite axion models with domain 
wall number $N_{\rm DW} = 1$. The key ingredient is the Lazarides–Shafi mechanism, which embeds the discrete symmetry 
responsible for domain walls into a continuous gauge symmetry, thereby formally reducing the 
domain wall number to one—an essential condition towards resolving the domain wall problem.%
\footnote{See also~\cite{Gherghetta:2025fip} for a composite axion model to avoid the domain wall problem with a bias term.}
Both of the examples presented predict cosmologically stable relics charged under QCD and electromagnetic interactions. However, the models can still be made phenomenologically viable 
by
introducing a secondary, low-scale mini-inflationary epoch that dilutes the abundance of heavy exotic states.
We identify a parameter space in which the axion string network re-enters the horizon before or around the QCD transition, allowing the dark matter abundance to be predominantly sourced by the collapse of string-wall systems—similar to the standard post-inflationary scenario.
This framework not only opens a generic cosmologically viable path for composite axions, but also highlights new model-building possibilities where high-quality Peccei–Quinn symmetries and realistic cosmology coexist.

The structure of the paper is as follows.  
In Sec.~\ref{sec:review}, we review the original composite axion model together with more recent developments, and provide a systematic derivation of the domain wall number.
In Sec.~\ref{sec:models}, we present two composite axion models with domain wall number equal to one. 
In Sec.~\ref{sec:cosmology}, we discuss the conditions required for a mini-inflation scenario that dilutes exotic heavy relics and realizes axion dark matter originating from the collapse of the string-wall network. 
Finally we summarize our main results and conclude.

\section{Review: Composite Axion Models and Domain Wall Problem}
\label{sec:review}

In this section, we review composite axion models, beginning with the original proposal by Kim and Choi~\cite{Kim:1984pt,Choi:1985cb}. These models aim to address the hierarchy between the PQ symmetry breaking scale and the Planck scale via strong dynamics. However, early constructions suffer from the axion quality problem, as the vector-like gauge theories employed allow PQ-violating operators consistent with gauge symmetries. To resolve this issue, later models introduce accidental PQ symmetries, requiring a chiral gauge structure to forbid explicit PQ-breaking terms. 
We also examine the domain wall number in these constructions, which faces the domain wall problem in post-inflationary scenarios.

\subsection{Vector-like Composite Axion: Original Model}
\label{sec:kim}

Consider an $SU(N)$ gauge theory with four pairs of vector-like fermions transforming in the fundamental and anti-fundamental representations of $SU(N)$. The theory is assumed to respect a global $SU(4)_L \times SU(4)_R$ flavor symmetry. We then gauge an $SU(3)$ subgroup of the diagonal subgroup $SU(4)_V \subset SU(4)_L \times SU(4)_R$, identifying it with the QCD color gauge group $SU(3)_c$.%
\footnote{Gauging the $SU(3)$ subgroup explicitly breaks the flavor symmetry, so the full $SU(4)_L\times SU(4)_R$ is no longer preserved, strictly speaking.}
The PQ charge assignment corresponds to the charge assignment of the $U(1)$ subgroup of $SU(4)$ flavor symmetry. The matter content is summarized in Table~\ref{tab:matter_contents_kim_choi}.

The $U(1)_{\rm PQ}$ symmetry charge assignment is also given in Table~\ref{tab:matter_contents_kim_choi}.%
\footnote{Other choices of the $U(1)_{\mathrm{PQ}}$ charge assignment are possible, which are equivalent to the one shown in Table~\ref{tab:matter_contents_kim_choi} up to global symmetry transformations. Such differences do not affect the following discussion.}
The $U(1)_{\rm PQ}$ symmetry satisfies the anomaly conditions,
\[
\mathcal{A}(U(1)_{\rm PQ} - SU(3)_c^2) = N, \quad 
\mathcal{A}(U(1)_{\rm PQ} - SU(N)^2) = 0.
\]
A nonzero anomaly coefficient $\mathcal{A}(U(1)_{\rm PQ} - SU(3)_c^2)$ is required to ensure that the axion couples to the $SU(3)_c$ topological term. Conversely, the PQ symmetry must be anomaly-free with respect to the $SU(N)$ gauge group, i.e.,
\(
\mathcal{A}(U(1)_{\rm PQ} - SU(N)^2) = 0,
\)
so that the axion does not acquire a potential from the strong dynamics of $SU(N)$, which would otherwise compromise the solution to the strong CP problem.

\begin{table}[h!]
    \centering
    \renewcommand{\arraystretch}{1.5}
    \begin{tabular}{|c|c|c|c|c|c|c|c|c|c|}
        \hline
        &  $[SU(3)_c]$  & $SU(4)_L^{\text{approx.}}$ & $[SU(N)]$  &  $SU(4)_R^{\text{approx.}}$  & $U(1)_{\rm PQ}$ &  $\mathbf{Z}_{N}$  \\ \hline
       $Q$&    $\mathbf{3}$ & \multirow{2}{*}{$\mathbf{4}$} & $N$ & & 1 & 1\\ 
       \cline{1-2} \cline{4-7}
       $\eta$ & $\mathbf{1}$ & & $N$ &  & -3 & 1 \\ \hline
         $\bar Q$ &  $\bar{\mathbf{3}}$ & & $\bar N$ & \multirow{2}{*}{$\bar{\mathbf{4}}$} & $0$ & $-1$\\ 
          \cline{1-4}\cline{6-7}
        $\bar\eta$ &   $\mathbf{1}$ & & $\bar N$ & & $0$ &  $-1$ \\ \hline
    \end{tabular}
    \caption{Matter content for the original composite axion model. $[G]$ denotes the gauge group $G$. 
    $SU(4)_{L,R}^{\text{approx.}}$ indicate the global symmetries of the model in the limit of zero $SU(3)_c$ gauge coupling.
    $\mathbf{Z}_{N}$ denotes the center symmetry of $SU(N)$.
    }
    \label{tab:matter_contents_kim_choi}
\end{table}

Below the confinement scale of $SU(N)$, the fermion condensates spontaneously break the flavor symmetry as
\[
    SU(4)_L \times SU(4)_R \to SU(4)_V
\]
in the limit of vanishing QCD coupling $g_c \to 0$. In this limit, there is no distinction between $Q$ and $\eta$; the global symmetry is fully $SU(4)_L \times SU(4)_R$, and the orientation of the condensates is completely arbitrary. That is, $\eta$ and the components of $Q$ can be freely mixed.
When the QCD coupling is nonzero, the condensates form so as to leave $SU(3)_c$ unbroken:
\begin{align}
   \langle Q \bar Q \rangle &\neq 0, \quad \langle \eta \bar \eta \rangle \neq 0, \\
   \langle Q \bar \eta \rangle &= \langle \eta \bar Q \rangle = 0.
\end{align}
A nonzero value of $\langle Q \bar \eta \rangle$ or $\langle \eta \bar Q \rangle$ would give mass to the $SU(3)_c$ gauge bosons, which is energetically disfavored.

The condensation leads to fifteen pseudo-Goldstone bosons associated with the breaking 
\(SU(4)_L \times SU(4)_R \to SU(4)_V\), including one gauge singlet boson and other \(SU(3)_c\)-charged bosons. 
The colored bosons become massive for the same reason that the charged pion is heavier than the neutral pion.
The remaining singlet boson corresponds to the axion.%
\footnote{In terms of the product
\[
    ({\bf 3} \oplus {\bf 1}) \otimes ({\bf \bar{3}} \oplus {\bf 1}) = {\bf 1} \oplus {\bf 1} \oplus {\bf 3} \oplus {\bf \bar{3}} \oplus {\bf 8},
\]
we have an additional singlet boson. However, this singlet can acquire a mass from the \(SU(N)\) strong dynamics; that is, the pseudo-Nambu-Goldstone boson corresponds to the \(SU(N)\)-anomalous \(U(1)\) symmetry.}

This model suffers from the quality problem. In particular, one could write down the explicit mass terms such as $Q\bar Q$ and $\eta\bar{\eta}$ allowed by the gauge symmetry in the lagrangian.
In other words, the PQ symmetry does not emerge as an accidental symmetry.
The suppression of these mass terms relies on the assumption of an approximate global $SU(4)_L \times SU(4)_R$ symmetry.

\subsubsection{Domain wall number from the master formula}
\label{sec:master-formula}
Let us calculate the domain wall number in the model presented above. Since it will be one of the main topics of the current paper we will present a generic formula for domain wall numbers following the recent discussion in the Ref\cite{Lu:2023ayc}. Let us suppose that the UV completion of our composite axion theory contains fermions $\psi_j$ with mutually prime integer  PQ charges $p_j$
\bea
&& {\rm gcd} (p_1,...,p_n)=1.
\eea
Then obviously the action must remain invariant under the transformation  
\bea
\label{eq:rotation}
&&\psi_j\to \psi _j\times \exp(i 2\pi p_j).
\eea
On the other hand, under such anomalous rotation the action changes as follows:
\bea
\label{eq:kappaG}
&&\int \frac{1}{8\pi^2}\theta {\rm tr}(G\wedge G)\to \int \frac{1}{8\pi^2}(\theta + 2\pi k_G ){\rm tr}(G\wedge G)\nn
&&k_G=\sum_i p_i\times  \text{dim} (r_i)\times 2 I(s_i),
\eea
where 
$\text{dim}(r_i)$ are the 
dimensions of the fermionic representations under the confining group and 
$I(s_i)$ is the corresponding Dynkin index under QCD normalized to $1/2$ for the fundamental 
representation. Thus it is obvious that we can identify
\bea
\theta\to \theta+2\pi k_G.
\eea
Since QCD potential is periodic in $\theta$ with a period $2\pi$,   it is  evident that the domain wall number is equal to $k_G$, as it counts the 
number of distinct QCD vacua encountered during a 
trivial rotation (Eq. 
\ref{eq:rotation}) equivalent to identity. However this 
discussion was not completely 
accurate because there could be a remaining gauge symmetry 
transformations making the counting above wrong, i.e. the center symmetry of the gauge group.
In UV theory these transformations correspond to (for example for $SU(N)$)
\bea
\psi_j \to \psi_j \exp\l(\frac{i 2\pi c_i}{N}\r),
\eea
where $c_i$ is N-ality of the representation.
On the other hand these transformations modify $\theta$ term as follows:
\bea
\label{eq:kappac}
&&\int \frac{1}{8\pi^2}\theta {\rm tr}(G\wedge G)\to \int \frac{1}{8\pi^2}(\theta + 2\pi k_c ){\rm tr}(G\wedge G)\nn
&&k_c=\sum_i \frac{c_i}{N}\times  {\rm dim} (r_i) 2 I(s_i).
\eea
Combining the Eq. (\ref{eq:kappaG})-(\ref{eq:kappac}) we can see that the domain wall number will be given by the greatest common divisor of $k_G,k_c$
\bea
\label{eq:master}
N_{\rm DW}={\rm gcd}(k_G,k_c).
\eea
Let us apply this discussion to the model above.
\bea
k_G=N,~~k_c=0~~\Rightarrow N_{\rm DW}={\rm gcd}(N,0)=N
\eea

In vector-like composite axion models, it is generally difficult--if not impossible--to simultaneously satisfy the following three conditions:
\begin{itemize}
    \item The PQ symmetry is (by definition) anomalous under \( SU(3)_c \)
    \item The PQ symmetry is non-anomalous under the confining non-abelian gauge group
    \item The domain wall number is \( N_{\rm DW} = 1 \)
\end{itemize}
This incompatibility motivates us to consider chiral gauge theories for composite axion models,%
\footnote{This point was already noted in the original paper~\cite{Choi:1985cb}.}
which we explore in the next subsection.

\subsection{Non-Vector-like Composite Axion: a moose-inspired extension of original model}
\label{sec:deconstruction}

\begin{table}[th]
    \centering
    \renewcommand{\arraystretch}{1.5}
    \begin{tabular}{|c|c|c|c|c|c|c|c|c|c|c|}
        \hline
        Field & $[SU(3)_0]$ & $SU(4)_L^{\text{approx.}}$ & $[SU(N)_1]$ & $[SU(4)_1]$ &  $[SU(N)_2]$ &  $SU(4)_R^{\text{approx.}}$ & $U(1)_{\rm PQ}$ &  $\mathbb{Z}_N$ \\ \hline
        $Q$ &  $\mathbf{3}$ &  \multirow{2}{*}{$\mathbf{4}$} & $\bar{N}$ &  & & & 1 & -1  \\ \cline{1-2} \cline{4-9}
        $\eta$ & $\mathbf{1}$ & & $\bar{N}$ &  & &  & -3 & -1 \\ 
        \hline
        $\chi_{1,1}$ &  &  & $N$ & $\bar{\mathbf{4}}$ &  &  & 0 & 1 \\ \hline
        $\psi_{1,2}$ &  &  & & $\mathbf{4}$ &  $\bar N$ & &  0 & 0  \\ \hline
         $\bar Q$ &  $\mathbf{\bar 3}$  &  & & & $N$ & \multirow{2}{*}{$\bar{\mathbf{4}}$} & 0 & 0  \\ 
         \cline{1-6}\cline{8-9}
        $\bar\eta$ & $\mathbf{1}$ & & &  & $N$ & & 0 & 0   \\ \hline
    \end{tabular}
    \caption{Matter contents. Extension of Choi-Kim. $\mathbf{Z}_N$ denotes the center symmetry of $SU(N)_1$.
    }
    \label{tab:matter_contents_ekc}
\end{table}

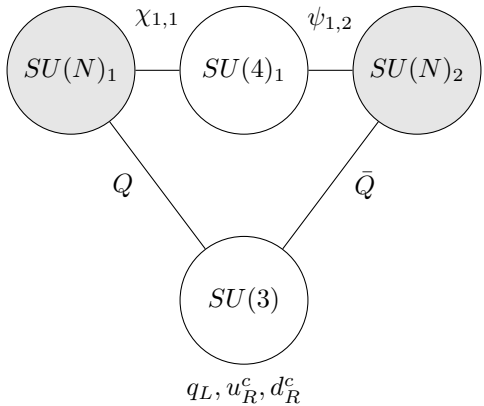
\begin{figure}[t] 
  \centering
\begin{tikzpicture}[xscale=2.3,yscale=1.7, 
every node/.style={font=\footnotesize}
]
  \node[circle, draw, fill=gray!20,  minimum size=1.7cm, inner sep=0pt] (N1) at (0,0) {$SU(N)_1$};
  \node[circle, draw,  minimum size=1.7cm, inner sep=0pt] (SU4_1) at (1,0) {$SU(4)_1$};
  \node[circle, draw,  fill=gray!20, minimum size=1.7cm, inner sep=0pt] (N2) at (2,0) {$SU(N)_2$};
  
  \node[circle, draw, minimum size=1.7cm, inner sep=0pt] (SU3) at (1,-1.8) {$SU(3)$};
  
  \node at (0.5,0.4) {$\chi_{1,1}$};
  \node at (1.5,0.4) {$\psi_{1,2}$};

  \draw[-] (N1) -- (SU4_1);
  \draw[-] (SU4_1) -- (N2);
  \draw[-] (N2) -- (SU3);

  \draw[-] (N1) -- (SU3);

  \node at (0.3,-0.9) {$Q$};
  \node at (1.7,-0.9) {$\bar{Q}$};
  \node at (1,-2.5) {$q_L, u^c_R, d^c_R$};
  
\end{tikzpicture}
\caption{Moose diagram of the accidental composite axion model.
 The bifundamental fermions $\chi_{1,1}$ and $\psi_{1,2}$ serve as link fields connecting adjacent gauge nodes. The SM quark fields reside in the $SU(3)$ gauge sector, which is connected to the chiral $SU(N)$ sector through the link fermions $Q$ and $\bar Q$.}
  \label{fig:moose_accidental}
\end{figure}

We consider chiral (non-vector-like) composite axion models constructed using Georgi’s moose diagram~\cite{Georgi:1985hf} or the deconstruction framework~\cite{Arkani-Hamed:2001kyx,Arkani-Hamed:2001nha}, following the concrete realization in~\cite{Redi:2016esr}. The gauge structure is
\[
SU(3)_0 \times SU(N)_1 \times SU(4)_1 \times SU(N)_2\ ,
\]
and the matter content is summarized in Table~\ref{tab:matter_contents_ekc}. The PQ charges are assigned so that the mixed anomalies \( \mathcal{A}(U(1)_{\mathrm{PQ}} - SU(N)_{1,2} - SU(N)_{1,2}) \) vanish. We adopt the PQ charge assignment given in Table~\ref{tab:matter_contents_ekc}.

This model can also be viewed as a combination of two vector-like composite axion models, where part of their flavor symmetries is gauged by an $SU(4)$. This realizes a chiral setup, in which direct fermion mass terms are forbidden—an important improvement over the original Kim--Choi model. 
In the limit of vanishing $SU(3)_0$ gauge coupling, the model possesses an accidental flavor symmetry
\[
SU(4)_L \times SU(4)_R\ .
\]  

We assume that the $SU(N)_1$ and $SU(N)_2$ gauge couplings are equal (and strong), while those of $SU(3)_0$ and $SU(4)_1$ are negligible. Under these assumptions, condensation of vector-like fermion pairs charged under $SU(N)_{1,2}$ breaks the flavor symmetry diagonally:
\[
SU(4)_L \times SU(4)_R \to SU(4)_V\ ,
\]
as in the original vector-like composite axion model. The unbroken $SU(3)$ subgroup corresponds to $SU(3)_c$, and the condensation also produces a gauge singlet boson—the axion.%
\footnote{The axion can acquire a potential from small instanton effects associated with the broken part of $SU(3)_0 \times SU(4)_1$~\cite{Aoki:2024usv}. The impact of the small instanton potential on the domain wall problem in composite axion models is left for future work.}

The same structure can be extended by adding more $SU(4)$ and $SU(N)$ gauge groups, and the model can be interpreted as a deconstruction of a 5D theory~\cite{Redi:2016esr}.

The quality problem still persists, although it is alleviated compared to the vector-like setup. In the current construction, the lowest-dimensional gauge-invariant operator that explicitly breaks the PQ symmetry is of dimension six, given by $\sim Q \chi_{1,1} \psi_{1,2} \bar Q$. The explicit breaking can be further suppressed by extending the moose diagram, inspired by the deconstruction framework. For example, introducing an additional set of sectors involving $SU(N)_3$ and $SU(4)_2$ raises the leading PQ-breaking operator to dimension nine, thereby resolving the quality problem for an axion decay constant of order $10^8$\,GeV.

Regarding the domain wall number, the center transformation does not lead to any further identification of the vacuum. 
Using a general formula, we find
\begin{align}
    k_G = N, \quad k_c = N\ ,
\end{align}
so that the domain wall number remains
\begin{align}
    N_{\mathrm{DW}} = \gcd(k_G, k_c) = N\ .
\end{align}
Here, $k_G$ and $k_c$ are computed for $SU(3)_c$ rather than $SU(3)_0$. 
In the next subsection, we consider a chiral composite axion model in which the center symmetry of the confining group partially identifies the vacua.

\subsection{Chiral Composite Axion: a chiral $SU(5)$ gauge theory}
\label{sec:automatic}
As an alternative to the theories  with vector-like confinement we can consider
models based on the chiral gauge theories. For example the {\it automatic Peccei-Quinn}  model~\cite{Gavela:2018paw}
is constructed using confinement and condensation in a chiral $SU(5)$ gauge theory.
\begin{table}[h!]
    \centering
    \renewcommand{\arraystretch}{1.5}
    \begin{tabular}{|c|c|c|c|c|c|c|c|c|}
        \hline
        Field & $[SU(5)]$ & $SU(3)_c$ & $U(1)_{\rm PQ}~(U(1)_B)$   \\ \hline
        $A$ &  $\mathbf{10}$ & $\mathbf{3}\oplus\bar{\mathbf{3}}$ & 1   \\ \hline
        $\bar F$ & $\mathbf{\bar 5}$ & $\mathbf{3}\oplus\bar{\mathbf{3}}$ & -3 \\ \hline
    \end{tabular}
    \caption{Automatic PQ symmetry model}
    \label{tab:automatic}
\end{table}

The field content of the model in Ref. \cite{Gavela:2018paw} consists of Weyl fermions in the \( \mathbf{10} \) and \( \overline{\mathbf{5}} \) representations. As is well known from the \( SU(5) \) GUT, the combination \( \mathbf{10} \oplus \overline{\mathbf{5}} \) forms a minimal anomaly-free set for a chiral, non-vector-like theory. These Weyl fermions are additionally charged under a real representation \( \mathbf{R} \) of the QCD color group \( SU(3)_c \), which allows all gauge anomalies involving both \( SU(5) \) and \( SU(3)_c \) to cancel. From the \( SU(5) \) point of view, the number of flavors of \( \mathbf{10} \oplus \overline{\mathbf{5}} \) is given by the dimension \( N_f \) of the real representation \( \mathbf{R} \) of \( SU(3)_c \). From the $SU(3)_c$ point of view, the reality of the representation ensures anomaly cancellation. In this work, we take \( \mathbf{R} = \mathbf{3} \oplus \overline{\mathbf{3}} \), while \( \mathbf{R} = \mathbf{8} \) is another simple possibility. The matter content is summarized in Table~\ref{tab:automatic}.

The PQ symmetry in this model arises as a \( SU(3)_c \)-anomalous but \( SU(5) \)-non-anomalous \( U(1) \) symmetry. From the viewpoint of the \( SU(5) \) gauge theory, the \( U(1)_{\text{PQ}} \) symmetry can also be regarded as a baryon number symmetry--i.e., a flavor-independent and non-anomalous \( U(1) \) symmetry.

A key question is the fate of the \( U(1)_{\text{PQ}} \) symmetry--specifically, whether it is spontaneously broken by the confinement dynamics of the \( SU(5) \) gauge theory. The Ref.~\cite{Gavela:2018paw} has assumed that the symmetry breaking occurs via the condensation of the operators
\bea
\langle\mathbf{10} \mathbf{10} \mathbf{10} \bar {\mathbf{5}}\rangle,~~
\langle \bar{\mathbf{5}}\bar{\mathbf{5}} \mathbf{10} \bar{\mathbf{5}}\bar{\mathbf{5}}\mathbf{10} \rangle \neq 0,
\eea
and indeed in this case  there is a spontaneous breaking of the PQ symmetry leading to the appearance of the composite axion in the IR spectrum of the theory. 
At the same time analyses based on 't~Hooft anomaly matching~\cite{Gavela:2018paw} suggest that the  \( U(1)_{\text{PQ}} = U(1)_B \) baryon symmetry may remain unbroken. More concretely, if we consider three anomaly conditions:
\[
\mathcal{A}(U(1)_{\text{PQ}}^3), \quad 
\mathcal{A}(U(1)_{\text{PQ}} - SU(3)_c^2), \quad 
\mathcal{A}(U(1)_{\text{PQ}} - \text{gravity}^2),
\]
then these high-energy anomalies can be matched by a massless composite baryon  of the form \( \mathbf{10} \, \overline{\mathbf{5}} \, \overline{\mathbf{5}} \), which carries PQ charge \( -5 \) and transforms in ${\bf R}$ of $SU(3)_c$. Hence, the scenario with an unbroken \( U(1)_{\text{PQ}} \) is also  consistent with anomaly matching.

Which scenario of the confinement is realized -with or without baryon (PQ) symmetry breaking- 
remains an open question under active investigation
(see $e.g.$~\cite{Bolognesi:2020mpe,Smith:2021vbf,Csaki:2021xhi,Karasik:2022gve} for recent studies on chiral symmetry breaking in chiral gauge theories). While these works shed light on possible symmetry-breaking patterns, the option of spontaneous baryon (PQ) symmetry breaking is not yet excluded. 
Interestingly, applying  the Most Attractive Channel (MAC)~\cite{Raby:1979my} analysis
to study the dynamics of confinement (see appendix \ref{app:MAC}), we find that contrary to the  assumptions of the paper  ~\cite{Gavela:2018paw},  the operators 
\bea
\langle \mathbf{10} \mathbf{10}\rangle,~~\langle \bar{\mathbf{5}} \mathbf{10}\rangle
\eea
will get vevs. 
Nevertheless, it should be emphasized that the results of the MAC analysis cannot be regarded as a proof that condensation pattern proposed in the 
Ref\cite{Gavela:2018paw} is incorrect. In the next section we will present a model with $N_{\rm DW}=1$ assuming the symmetry breaking proceeds as was advocated in the Ref.\cite{Gavela:2018paw}.

At last we calculate the domain wall number in the model. It  is equal to $2$, as was already noted in the original work~\cite{Gavela:2018paw}. Indeed, the naive counting leads to  $N_{\rm DW}=10$. 
However, the center symmetry of $SU(5)$ also acts on the $\theta$-term of $SU(3)_c$, which modifies the result. 
Applying the general formula, we obtain
\bea
&& k_G=-3 \times 5 \times 2 + 1 \times 10 \times 2 = -10, \quad k_c=6, \nn
&&N_{\rm DW}=\gcd(k_c,k_G)=2 \ .
\eea
Thus, the domain wall problem persists in this model. 
The origin of $N_{\rm DW}=2$ is related to the fermions being in the  real representation of $SU(3)_c$, where the later condition was needed in order to cancel $SU(3)_c$ anomalies.

\section{Composite Axion Models with $N_{\rm DW}=1$}
\label{sec:models}

We present two composite axion models that realize a domain wall number 
$N_{\rm DW}=1$ through identification by the center symmetry of the continuous gauge groups. 
In our models, the composite axion arises from a chiral gauge theory, and the quarks charged under $SU(3)_c$ with non-trivial PQ charge are in the (anti-)fundamental 
representation of $SU(3)_c$, which appears to play an important role in realizing $N_{\rm DW}=1$.

\subsection{Composite Axion from Chiral Moose}
\label{sec:model1}
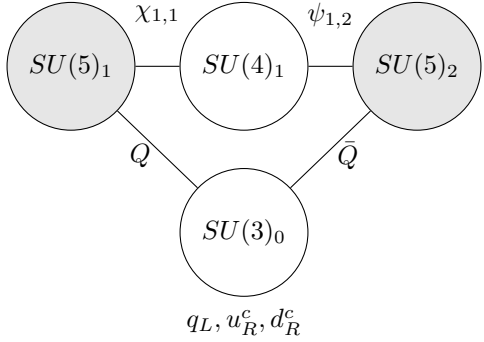
\begin{figure}[t] 
  \centering
\begin{tikzpicture}[xscale=2.3,yscale=1.7, 
every node/.style={font=\footnotesize}
]
  \node[circle, draw, fill=gray!20,  minimum size=1.7cm, inner sep=0pt] (N1) at (0,0) {$SU(5)_1$};
  \node[circle, draw,  minimum size=1.7cm, inner sep=0pt] (SU4_1) at (1,0) {$SU(4)_1$};
  \node[circle, draw,  fill=gray!20, minimum size=1.7cm, inner sep=0pt] (N2) at (2,0) {$SU(5)_2$};
  \node[circle, draw, minimum size=1.7cm, inner sep=0pt] (SU3) at (1,-1.3) {$SU(3)_0$};
  \node at (0.5,0.4) {$\chi_{1,1}$};
  \node at (1.5,0.4) {$\psi_{1,2}$};

  \draw[-] (N1) -- (SU4_1);
  \draw[-] (SU4_1) -- (N2);

  \draw[-] (N1) -- (SU3);
  \draw[-] (N2) -- (SU3);

  \node at (0.4,-0.7) {$Q$};
  \node at (1.6,-0.7) {$\bar{Q}$};
  \node at (1,-2) {$q_L, u^c_R, d^c_R$};
  
\end{tikzpicture}
\caption{Moose diagram of the chiral moose composite model.
 The fermions $\chi_{1,1}$ and $\psi_{1,2}$ serve as link fields connecting adjacent gauge nodes. The SM quark fields reside in the $SU(3)$ gauge sector, which is connected to the chiral $SU(5)$ sector through the link fermions $Q$ and $\bar Q$.}
  \label{fig:moose_chiral}
\end{figure}

We proceed by constructing a composite axion model based on $SU(5)$ chiral gauge theory extending the automatic axion proposal of the Ref.\cite{Gavela:2018paw},
in order to obtain $N_{\rm DW}=1$. The moose diagram representation of the model is shown in Fig.~\ref{fig:moose_chiral}. 
It is similar to the vector-like models
discussed in Section~\ref{sec:deconstruction}, except that  the confining $SU(N)$ groups are replaced by the chiral $SU(5)$ dynamics.
This model achieves $N_{\rm DW}=1$, avoids the Landau pole problem, and alleviates the axion quality problem.

\begin{table}[h!]
    \centering
    \renewcommand{\arraystretch}{1.5}
    \begin{tabular}{|c|c|c|c|c|c|c|c|}
        \hline
        Field & $[SU(3)_0]$ & $SU(4)_L^{\text{approx.}}$ & $[SU(5)_1]$ & \textbf{$[SU(4)_1]$} &  $[SU(5)_2]$ &  $SU(4)_R^{\text{approx.}}$ & $U(1)_{\rm PQ}$ \\ \hline
        $Q^\alpha_A$ &  $\mathbf{3}$ & \multirow{2}{*}{$\mathbf{4}$} & $\bar{\mathbf{5}}$ &  & & & -3   \\ 
        \cline{1-2} \cline{4-8}
        $\eta_A$ & $\mathbf{1}$ & & $\bar{\mathbf{5}}$ &  &  & & -3  \\ \hline
        $\chi_{1,1}^{AB\,\hat\alpha}$ &  & & $\mathbf{10}$ & $\mathbf{4}$ &  & & 1  \\ \hline
        $\psi_{1,2\,\hat\alpha}^{\hat A\hat B}$ &  & & & $\mathbf{\bar 4}$ &  $\mathbf{10}$ & & 0  \\ \hline
         $\bar Q_{\alpha\hat A}$ &  $\mathbf{\bar 3}$  &  & &  & $\mathbf{\bar 5}$ &  \multirow{2}{*}{$\bar{\mathbf{4}}$} & 0    \\ \cline{1-6}\cline{8-8}
        $\bar\eta_{\hat A}$ & $\mathbf{1}$ &  &  & & $\bar{\mathbf{5}}$ & & 0  \\ \hline
    \end{tabular}
    \caption{Matter contents. $\alpha$ denotes $SU(3)_0$ indices. $A,B$ denotes the $SU(5)_1$ indices. $\hat A,\hat B$ denote the $SU(5)_2$ indices, and $\hat\alpha$ denotes the $SU(4)_1$ indices. A gauge symmetry $G$ is denoted with $[G]$. 
    $SU(4)_{L,R}^{\text{approx.}}$ denote the global symmetries in the limit of zero $[SU(3)_0]$ gauge coupling.}
    \label{tab:matter_contents}
\end{table}

Concretely, the model includes two confining \( SU(5) \) gauge groups, denoted \( SU(5)_1 \) and \( SU(5)_2 \), each contains four flavors of chiral fermions in the \( \mathbf{\bar{5}} \oplus \mathbf{10} \) representation.
Each confining sector exhibits an (approximate) flavor symmetry \( SU(4)_{\bar{5}} \times SU(4)_{10} \), where the subscripts indicate the type of fermion associated with each flavor symmetry. We gauge a diagonal subgroup of the two \( SU(4)_{10} \) flavor symmetries, which we denote by \( [SU(4)_1] \), and likewise gauge a diagonal \( SU(3) \) subgroup of the two \( SU(4)_{\bar{5}} \) symmetries, denoted \( [SU(3)_0] \).
The matter content of the model is summarized in Table~\ref{tab:matter_contents}. To help guide the reader, we adopt the convention that gauged symmetries are enclosed in brackets, \([G]\), while global symmetries are left unbracketed, as before.
The $[SU(3)_0]$ and $[SU(4)_1]$ are weakly gauged sectors.
As discussed in more detail later, we assume that the condensates in the \( SU(5)_{1,2} \) sectors break 
\[
SU(3)_0 \times SU(4)_1 \rightarrow SU(3),
\]
where the unbroken \( SU(3) \) is identified with the QCD color group \( SU(3)_c \).

The PQ symmetry corresponds to the following rotation of the fermion fields:
\begin{align}
    Q \to Q\, e^{-3i\alpha}, \quad  
    \eta \to \eta\, e^{-3i\alpha}, \quad  
    \chi_{1,1} \to \chi_{1,1}\, e^{i\alpha},
\end{align}
as shown in Table~\ref{tab:matter_contents}, and is denoted by \( U(1)_{\rm PQ} \). This symmetry is anomaly-free with respect to the \( SU(5)_1 \) gauge group,
\begin{align}
    A(U(1)_{\rm PQ} - SU(5)_1 - SU(5)_1) = 0,
\end{align}
but is anomalous under QCD:
\begin{align}
    A(U(1)_{\rm PQ} - SU(3)_c - SU(3)_c) \neq 0.
\end{align}
To see the latter in more detail, we examine the transformation of the $\theta$-angles in the \( SU(3)_0 \) and \( SU(4)_1 \) sectors under the PQ rotation. The $\theta$-angles, denoted \( \theta_3 \) and \( \theta_4 \), shift as
\begin{align}
    \theta_3 \to \theta_3 - 3\alpha \times 5, \qquad
    \theta_4 \to \theta_4 + \alpha \times 10.
\end{align}
Here, the $\theta$-terms are defined by
\begin{align}
    S \supset \int \left(
    g_3^2 \frac{\theta_3}{8\pi^2} \mathrm{Tr}(F_3 \wedge F_3) +
    g_4^2 \frac{\theta_4}{8\pi^2} \mathrm{Tr}(F_4 \wedge F_4)
    \right),
\end{align}
where \( F_3 \) and \( F_4 \) are the field strengths of \( SU(3)_0 \) and \( SU(4)_1 \), respectively, and \( g_3 \), \( g_4 \) are the corresponding gauge couplings. We are working in the basis where the gauge kinetic terms are canonically normalized. After the condensation that breaks \( SU(3)_0 \times SU(4)_1 \to SU(3)_c \), the effective $\theta$-angle of the $SU(3)_c$ is given by
\begin{align}
    \theta_c = \theta_3 + \theta_4.
\end{align}
The effective theory below the condensation scale contains a single gauge field \( F \) for \( SU(3)_c \), and the original contributions from \( SU(3)_0 \) and \( SU(4)_1 \) can be matched as
\begin{align}
    g_3^2 \frac{\theta_3}{8\pi^2} \mathrm{Tr}(F_3 \wedge F_3) 
    &\sim g_c^2 \frac{\theta_3}{8\pi^2} \mathrm{Tr}(F \wedge F), \\
    g_4^2 \frac{\theta_4}{8\pi^2} \mathrm{Tr}(F_4 \wedge F_4) 
    &\sim g_c^2 \frac{\theta_4}{8\pi^2} \mathrm{Tr}(F \wedge F),
\end{align}
where \( F \) is the field strength of \( SU(3)_c \), and \( g_c \) is its effective gauge coupling, given by
\begin{align}
    \frac{1}{g_c^2} = \frac{1}{g_3^2} + \frac{1}{g_4^2}.
\end{align}
The matching of the gauge fields is ensured by the following rotation:
\begin{align}
    \begin{pmatrix}
        F_H \\
        F
    \end{pmatrix}
    =
    \frac{1}{\sqrt{g_3^2 + g_4^2}} 
    \begin{pmatrix}
        g_3 & g_4 \\
        -g_4 & g_3
    \end{pmatrix}
    \begin{pmatrix}
        F_3 \\
        F_4
    \end{pmatrix},
\end{align}
where \( F_H \) denotes the heavy gauge boson field strength that is integrated out after the symmetry breaking. Therefore, the PQ rotation induces a shift of the effective QCD theta angle as
\begin{align}
    \theta_c \to \theta_c - 5\alpha,
\end{align}
demonstrating that the PQ symmetry is indeed anomalous under QCD. One may also find a potential candidate for the PQ symmetry where $Q$ has the PQ charge $1$, and $\chi_{1,1}$ has the charge $-3$ because this symmetry is also non-anomalous with $SU(5)_{1,2}$ but anomalous with $SU(3)_c$. However, this charge assignment is not convenient to discuss the PQ symmetry breaking. See the appendix~\ref{app:more_on} for more detailed discussion about this point.

\subsubsection{Spontaneous symmetry breaking and condensates}

Let us discuss the spontaneous breaking of the flavor symmetries and subsequently the PQ symmetry via fermion condensations. We will assume that it 
proceeds very similar to the original model in Ref \cite{Gavela:2018paw} and reviewed in the section \ref{sec:automatic}.
A lowest-dimensional condensation operator that is gauge invariant under $SU(5)_1$ and breaks the chiral symmetries composed of $SU(3)_0$ and $SU(4)_1$ is
\begin{align}
  \chi_{1,1}\chi_{1,1} \chi_{1,1} Q \sim \mathbf{10} \, \mathbf{10} \, \mathbf{10} \, \bar{\mathbf{5}} \ ,
\end{align}
where on the right-hand side, we indicate the representation under $SU(5)_1$. We assume this condensation breaks the symmetry as
\begin{align}
 \langle  \chi_{1,1}  \chi_{1,1}  \chi_{1,1} Q \rangle \quad \Rightarrow \quad
  SU(3)_0 \times SU(4)_1 \to SU(3)_c \ .
\end{align}
Indeed, the operator $\chi_{1,1}  \chi_{1,1}  \chi_{1,1} Q$ contains a component transforming as a $\mathbf{3}$ of $SU(3)_0$ and a $\bar{\mathbf{4}}$ of $SU(4)_1$, given by
\begin{align}
\epsilon_{\hat\alpha\hat\beta\hat\gamma\hat\delta}\, \chi_{1,1}^{\hat\beta} \chi_{1,1}^{\hat\gamma} \chi_{1,1}^{\hat\delta} Q^\alpha \sim \Phi^\alpha_{1\,\hat \alpha} \ ,
\end{align}
where only the $SU(3)_0$ index $\alpha$ and the $SU(4)_1$ index $\hat\alpha$ are shown. We define $\Phi_1 \sim\chi_{1,1} \chi_{1,1} \chi_{1,1} Q$ as the composite field triggering the symmetry breaking.
We can perform a similar analysis for the $SU(5)_2$ sector. The condensation there involves the operator
\begin{align}
    \langle \psi_{1,2} \psi_{1,2} \psi_{1,2} \bar{Q} \rangle \neq 0 \ .
\end{align}
Importantly, this condensation does \emph{not} break the $U(1)_{\rm PQ}$ symmetry, since the operator is invariant under the PQ rotation.

The lowest-dimensional gauge-invariant operators that carry nonzero $U(1)_{\rm PQ}$ charge are
\begin{align}
    Q^2\chi_{1,1} \psi_{1,2} \bar Q^2,
    \quad Q\eta\chi_{1,1} \psi_{1,2} \bar Q\bar\eta,
     \quad  \eta^2\chi_{1,1} \psi_{1,2}\bar\eta^2,
    \sim ({\bf 10}{\bf \bar 5}{\bf \bar 5})_{SU(5)_1} ({\bf 10}{\bf \bar 5}{\bf \bar 5})_{SU(5)_2} \ .
\end{align}
Here, the right-hand side indicates the representations under $SU(5)_{1,2}$. Each of these operators carries a total $U(1)_{\rm PQ}$ charge of $-5$.
We assume that the strong dynamics of the $SU(5)_{1,2}$ sectors lead to the condensation of such operators,
\begin{align}
    \left\langle ({\bf 10}{\bf \bar 5}{\bf \bar 5})_{SU(5)_1} ({\bf 10}{\bf \bar 5}{\bf \bar 5})_{SU(5)_2} \right\rangle \neq 0 
    \quad \Rightarrow \quad \text{SSB of}~~U(1)_{\rm PQ} \ .
\end{align}
As discussed in Sec.~\ref{sec:automatic} this pattern of condensation is not supported by the MAC analysis (see appendix~\ref{app:MAC}) but we do not consider it as a no-go.
We continue by assuming that the $U(1)_{\rm PQ}$ symmetry is indeed spontaneously broken triggered by the confining dynamics.

Let us also discuss the axion quality problem.  
A lowest-dimensional Planck-suppressed operator that explicitly breaks the $U(1)_{\rm PQ}$ symmetry is given by%
\footnote{Investigation of small instanton effects is left for future research.}
\begin{align}
  \sim  \frac{  Q^2\chi_{1,1} \psi_{1,2} \bar Q^2}{M_P^5} \sim \frac{\Lambda^9}{M_P^5}\ ,
\end{align}
where $M_P$ is the reduced Planck mass, and $\Lambda$ denotes the dynamical scale of the $SU(5)_{1,2}$ gauge sectors.
To avoid spoiling the axion solution to the strong CP problem, the size of this PQ-violating term must be much smaller than the QCD contribution to the axion potential:
\begin{align}
    \frac{\Lambda^9}{M_P^5} \lesssim 10^{-10} \Lambda_{\rm QCD}^4\ .
\end{align}
For a reference scale $\Lambda \sim 10^{10}\, \mathrm{GeV}$, this condition is not satisfied and the quality problem persists.  
Nevertheless, the required fine-tuning of parameters is drastically relaxed compared to conventional non-composite axion models.
Moreover, if we consider a lower axion decay constant, such as $F_a \sim 10^8\, \mathrm{GeV}$, the PQ-violating contribution can remain sufficiently suppressed.
Finally, by adding two more sites to the moose diagram with an additional $SU(5)_{3,4}$ chiral gauge group, the minimal dimension of the gauge-invariant operator can be further raised to 14 (see Appendix~\ref{app:more_on} for an extended version of the moose diagram).

\subsubsection{Domain wall number}

We can calculate the domain wall number using our master formulas also in this case.
Since $SU(3),SU(4)$ are broken to diagonal $SU(3)_{QCD}$, we will get
\bea
k_G^{QCD}=k_G^{SU(3)}+k_G^{SU(4)}=-3\times 5+10=-5.
\eea
For the calculation of $k_c$ we 
 note that there will be two center transformations which remain unbroken:
\bea
&&k_c^{SU(3)}|_{SU(5)_1}=-1,~~~~
k_c^{SU(4)}|_{SU(5)_1}=4,\nn
&&k_c^{SU(3)}|_{SU(5)_2}=-1,~~~~
k_c^{SU(4)}|_{SU(5)_2}=4,\nn
\eea
so that in total 
\bea
k_c|_{SU(5)_{1,2}}=3,
\eea
thus the domain wall number will be
\bea
N_{\rm DW}=\gcd(5,3)=1.
\eea
The equation above shows that there exist string solutions with $N_{\rm DW}=1$. Establishing whether during the cosmological evolution only this type of strings remains  
requires numerical simulations,
but  establishing $N_{\rm DW} = 1$ is  an important milestone toward a viable post-inflationary composite axion model.

\subsection{Composite Axion from a ``Gauged'' $U(1)$ Peccei-Quinn Symmetry}
\label{sec:gauged_pq}
In this section, we present an alternative realization of composite axion scenario  with $N_{\rm DW}=1$. The model can be 
interpreted as a composite version of a ``gauged'' $U(1)$ Peccei–Quinn symmetry ideas proposed in \cite{Barr:1992qq,Fukuda:2017ylt}. 
It consists of two well-separated sectors, each based on the original Kim-Choi composite axion model. 
Both sectors are vector-like $SU(N)$ gauge theories, and they are coupled  only 
through an additional $U(1)$ gauge symmetry under which the $SU(N)$ vector-like quarks are charged.
As a result, the full theory becomes chiral, and vector-like 
mass terms are forbidden. This chiral $U(1)$ gauge symmetry 
also plays a crucial role in identifying vacua, thereby 
ensuring that the domain wall number is $N_{\rm DW} = 1$.\\
\begin{table}[h!]
    \centering
    \renewcommand{\arraystretch}{1.5}
    \begin{tabular}{|c|c|c|c|c|c|c|c|}
        \hline
        &  $SU(3)_c$ & $SU(N)$ & $U(1)_{\rm PQ_1}$ &  $U(1)_{\rm gPQ}$  \\ \hline
       $Q$&    $\mathbf{3}$ & $N$ & 1 & 2M\\ \hline
       $\eta$ & $\mathbf{1}$ & $N$ & -3 & 7M \\ \hline
         $\bar Q$ &  $\bar{\mathbf{3}}$ &  $\bar N$ & $0$ & $-5M$\\ \hline
        $\bar\eta$ &   $\mathbf{1}$ & $\bar N$ & $0$ &  $2M$ \\ \hline
    \end{tabular}
     \begin{tabular}{|c|c|c|c|c|c|c|c|}
        \hline
        &  $SU(3)_c$ & $SU(M)$ & $U(1)_{\rm PQ_2}$ &  $U(1)_{\rm gPQ}$   \\ \hline
       $Q'$&    $\mathbf{3}$ & $M$ & 1 & -2N\\ \hline
       $\eta'$ & $\mathbf{1}$ & $M$ & -3 & -7N\\ \hline
         $\bar Q'$ &  $\bar{\mathbf{3}}$ &  $\bar M$ & $0$ & $5N$\\ \hline
        $\bar\eta'$ &   $\mathbf{1}$ & $\bar M$ & $0$ &  $-2N$ \\ \hline
    \end{tabular}
    \caption{Matter contents. }
    \label{tab:gauged_pq}
\end{table}

More concretely,  the first sector consists of an  $SU(N)$ gauge theory with four flavors of vector-like fermions in the (anti-)fundamental representations of $SU(N)$. A diagonal $SU(3)$ subgroup of the flavor symmetry group $SU(4)_L \times SU(4)_R$ is gauged and is identified with QCD color group, $SU(3)_c$. 
The second sector has the same structure, except that the confining group is  $SU(M)$ with $N\neq M$.
Each of the sectors possesses its own $U(1)_{\rm PQ_{1,2}}$ symmetry which is spontaneously broken during  the confinements of the $SU(N), SU(M)$ groups. Remember that $U(1)_{\rm PQ_{1,2}}$ are defined only up to the addition of unbroken $U(1)$ transformations and in  our case
there are four unbroken vector-like flavor symmetries 
$U(1)_Q\times U(1)_\eta\times U(1)_{Q'}\times U(1)_{\eta'}$. 
These symmetries are defined as follows:
under $U(1)_a$ only the fields  $a,\bar a$ are charged and  have opposite charges and the rest  are neutral. 
Taking this into account one can show that there is a particular linear combination of $U(1)_{\rm PQ_{1,2}}$ which is anomaly free under $SU(3)_c$, when both sectors are included. The charges for this linear combination $U(1)_{\rm gPQ}$ are listed in the Table \ref{tab:gauged_pq}.
This choice of the charges  remarkably cancels all unwanted anomalies involving $\mathcal{A}(U(1)_{\rm gPQ}^3)$ using only the fermions listed in Tab.~\ref{tab:gauged_pq}.
Concretely, with the charge assignment, the $U(1)_{\rm gPQ}$ symmetry is anomaly-free within {\it each sector}.\footnote{Our charges follow the results presented  on the  Table 1 in~\cite{Batra:2005rh}. Another possible charge assignment would be $(-4,1,5,4)$ for the fields ($Q,\bar Q,\eta,\bar \eta$).
}

\begin{align}
    &\mathcal{A}(U(1)_{\rm gPQ} - SU(N)^2) = 0\ , \\
    &\mathcal{A}(U(1)_{\rm gPQ} - SU(M)^2) = 0\ , \\
    &\mathcal{A}(U(1)_{\rm gPQ}^3) = 0\ , \\
    &\mathcal{A}(U(1)_{\rm gPQ} - \text{gravity}^2) = 0\ .
\end{align}
On the other hand, the anomaly with respect to $SU(3)_c$ cancels only when both sectors are taken into account:
\begin{align}
    \mathcal{A}(U(1)_{\rm gPQ} - [SU(3)_c]^2) \propto (2 - 5)MN - (2 - 5)MN = 0\ .
\end{align}
As mentioned earlier, 
due to the $U(1)_{\rm gPQ}$ gauge symmetry, vector-like mass terms for the quarks are forbidden, and the resulting theory is chiral with respect to this $U(1)$.

As in the Kim–Choi model, below the confinement scales of \(SU(N)\) and \(SU(M)\), fermion condensation spontaneously breaks the flavor symmetries as
\[
    SU(4)_L \times SU(4)_R \to SU(4)_V, \quad SU(4)_L' \times SU(4)_R' \to SU(4)_V',
\]
where the prime denotes the flavor symmetries associated with the \(SU(M)\) sector.
The condensates take the form
\begin{align}
    &\langle Q\bar{Q} \rangle\neq 0,  \quad
     \langle \eta \bar{\eta} \rangle \neq 0, \quad
    \langle Q \bar{\eta} \rangle = \langle \eta \bar{Q} \rangle = 0, \\
    &\langle Q'\bar{Q}' \rangle \neq 0,\quad \langle \eta' \bar{\eta}' \rangle \neq 0, \quad
    \langle Q' \bar{\eta}' \rangle = \langle \eta' \bar{Q}' \rangle = 0.
\end{align}
The condensates with 
 \(\langle \eta \bar{Q} \rangle \neq 0\) or \(\langle Q \bar{\eta} \rangle \neq 0\),  are energetically disfavored, since they lead to the \(SU(3)_c\) gauge fields becoming massive.

Let us identify the degree of freedom corresponding to the  composite axion. It must be invariant under the under $U(1)_{\rm gPQ}$ transformations.
We can find it by looking at the degrees of freedom contained in the phases of the condensates:
\begin{align}
    \langle Q \bar Q\rangle \sim e^{i\theta_1}, \qquad \langle Q' \bar Q'\rangle \sim e^{-i\theta_2}.
\end{align}
Under $U(1)_{\rm gPQ}$
\bea
&&\theta_1\to \theta_1-3 M \theta_{\rm gPQ},~~~\theta_2\to \theta_2-3 N \theta_{\rm gPQ}\Rightarrow\nn
&&\theta= N \theta_1-M \theta_2 ~~{\rm gauge~ invariant}\Rightarrow {\rm axion}.
\eea
Exactly the same field will appear in the anomaly term for the interactions with gluons,
\begin{align}
    \mathcal{S}= \int \frac{ \left[\left(N \theta_1-M\theta_2\right)=\theta\right]}{8\pi^2}\text{tr}(G\wedge G)\ .
\end{align}
One can see it by removing the phases $\theta_{1,2}$ from the condensates by  the chiral rotation of the $Q,Q'$ fields. This highlights once more that the physical axion is equal to:
\begin{align}
\label{eq:physical_axion}
 \theta=   N \theta_1-M\theta_2.
\end{align}

Exactly the same combination of the phases appears also in the lowest gauge invariant operator that breaks the global PQ symmetry explicitly,
\begin{align}
    \frac{(Q\bar Q)^N (Q'\bar Q')^M}{M_P^{3N+3M-4}}\ .
\end{align}

The axion quality problem is addressed in the benchmark scenario with \( N = 3 \) and \( M = 4 \), where the leading Planck-suppressed PQ-violating operator takes the form  
\begin{align}
      \sim \frac{(Q\bar Q)^3 (Q'\bar Q')^4}{M_P^{17}}\ .
\end{align}  
This high-dimensional operator provides sufficient suppression to protect the axion potential for our interested parameter range in $F_a<10^{11-12}$\,GeV.

\begin{figure}[htbp]
    \centering
    \includegraphics[width=0.5\textwidth]{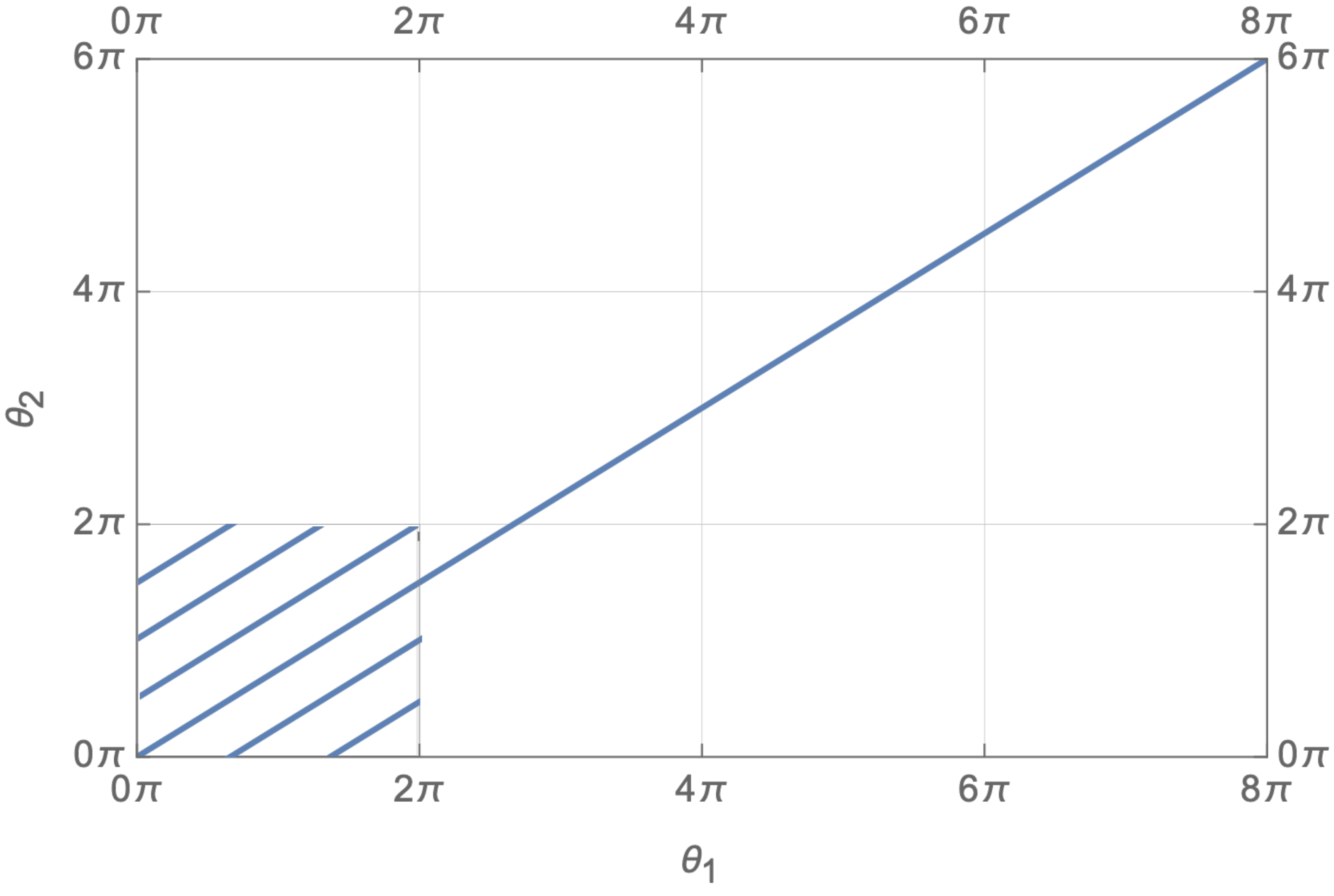}
    \caption{ Physical domain for $N=3$ and $M=4$.  }
       \label{fig:domain}
\end{figure}

Let us look now at the domain wall number in the model. The angles $\theta_{1,2}$ are defined in the range
\begin{align}
    \theta_1, \theta_2 \in [0,2\pi).
\end{align}
At the same time the gauge equivalent trajectory is $\theta_2=N \theta_1/M$. This is illustrated on the  
Fig.~\ref{fig:domain}, drawn for the case \(N = 3\), \(M = 4\), where the blue line represents the gauge-equivalent trajectory \(\theta_2 = \frac{3}{4} \theta_1\). This implies that the physical domain is restricted to \(\theta_1 \in [0, 2\pi/3]\) and \(\theta_2 \in [0, 2\pi/4]\), and therefore the domain wall number is
\begin{align}
    N_{\rm DW} = 1\ .
\end{align}

We can also obtain the domain wall number using our master formula. 
Let us suppose that  the confining scales for two groups $SU(M)$ and $SU(N)$ are different, and for concreteness we will assume that $\Lambda_{SU(M)}\gg \Lambda_{SU(N)}$. Then the symmetry breaking will proceed as follows. First, $SU(M)$ will confine, breaking  the gauge symmetry down to
\bea
U(1)_{\rm gPQ}\to Z_{3N},
\eea
since the chiral condensates 
$\langle Q' \bar Q'\rangle$ and $\langle \eta' \bar\eta'\rangle$
have $U(1)_{\rm gPQ}$ charges  $3N$ and $-9N$
respectively.
Under  this remaining $Z_{3N}$  
all of the fields will 
transform as follows,
\bea
\Psi\to e^{\frac{2\pi i}{3N} Q_{gPQ}}\Psi.
\eea
At this point we can look at the symmetries of the theory in IR. On top of the remaining center symmetry of the  $SU(N)$ group  there will be an additional gauge $Z_{3N}$ remnant of $U(1)_{\rm gPQ}$, which can  identify some of the QCD vacua. This can be taken into account similarly to the section \ref{sec:master-formula} by introducing $k_{Z_{3N}}^{SU(N)}$. Then, we find :
\bea
&&k_G^{SU(N)}=N, k_c^{SU(N)}=0\nn
&&k_{Z_{3N}}^{SU(N)}=\left(\frac{2M}{3N}-\frac{5M}{3N}\right)\times N\times 1=M\nn
&&N_{\rm DW}={\rm gcd}\left(k_G^{SU(N)},k_c^{SU(N)},k_{Z_{3N}}^{SU(N)}\right)={\rm gcd}(N, 0,M)={\rm gcd}(N,M)=1.
\eea

This condition similarly to the discussion in the section \ref{sec:model1} serves only as a proof of existence of $N_{\rm DW}=1$ strings. Whether or not this type of strings will dominate the universe can be determined only through the dedicated numerical simulations. Similar setups for the fundamental scalars have been studied in~\cite{Hiramatsu:2019tua,Hiramatsu:2020zlp}; however, those results cannot be directly 
extrapolated to our case, mainly due to the different choices of 
$N$ and $M$. To conclusively determine whether the present setup resolves the domain wall 
problem, more simulations will be needed.

Several other issues arise in this setup. 
The first issue is the Landau pole problem, particularly for the $U(1)_{\rm gPQ}$ gauge coupling, 
which inevitably develops a Landau pole below the Planck mass scale unless the $U(1)_{\rm gPQ}$ gauge coupling is sufficiently small, 
due to the presence of additional matter fields.%
\footnote{The size of the $U(1)_{\rm gPQ}$ gauge coupling affects the types of cosmic strings that are formed.}
We also discussed the asymptotic freedom condition of $SU(3)_c$ in Appendix~\ref{app:beta}, which requires $N+M \leq 10$.
The second issue concerns the presence of heavy particles, i.e. mesons or baryons from the $SU(M)$ and $SU(N)$ strong dynamics, charged under $SU(3)_c$.
These particles are in thermal equilibrium with the standard model plasma in the early universe. After the QCD transition, they can bind with Standard Model quarks, forming stable or long-lived composite states with non-zero (fractional) electromagnetic charge.
Such fractionally charged heavy relics are tightly constrained by a variety of observations and cannot constitute the dominant component of dark matter. We will discuss this cosmological issue in detail in the next section.

\section{Cosmological Challenges and Possible Resolutions}
\label{sec:cosmology}
In the previous sections we have been focusing on finding the composite axion models satisfying the $N_{\rm DW}=1$ condition.
In such models, the cosmological evolution may be such that only strings with $N_{\rm DW} \leq 1$  remain, thereby resolving the domain wall problem.
However, in addition to this 
issue, composite models face 
another challenge: the 
existence of charged, stable particles that are thermally produced in the early universe.
These states are 
baryons and mesons that appear after the confinement of (dark) non-Abelian gauge sectors  and the breaking of  PQ symmetry.
 Following the QCD  transition, they can bind to SM quarks, forming charged massive particles (CHAMPs).
These states acquire masses of order the confinement scale,%
\footnote{Meson masses are generated with a loop suppression.} and
their  relic abundance  is very strongly  constrained (see for example \cite{Dunsky:2018mqs} for a recent analysis). We are aware of two 
possibilities for solving the problem with overabundance of charged stable  particles: (i) enabling their decay into the Standard Model sector, and (ii) diluting their abundance via second mini-inflation. 
The former can be realized by assigning appropriate \( U(1)_{\text{EM}} \) electro-magnetic charges to the fermions. However, this generally requires introducing additional fields to cancel gauge anomalies. These extra fields, in turn, may introduce new cosmological issues, such as further stable relics.
In this work, instead, we mainly focus on the latter approach of dilution, which will be applicable to a broad class of axion models with dangerous relics.
Specifically, we consider a scenario in which a second inflationary epoch dilutes the heavy relics, while the string network reenters the horizon before or even after the QCD transition, and the string-wall network eventually collapses due to $N_{\rm DW}=1$. In this way, the axion dark matter abundance is dominated by the contribution from topological defects as the post-inflationary scenario.

\subsection{Diluting Relics}
We now estimate the baryon energy density after dilution by a brief period of inflation, which we refer to as the “second inflation”.
We will assume that the baryon number density is set by thermal freeze-out, with a thermally averaged annihilation cross section
\begin{align}
\langle \sigma v \rangle \sim \frac{1}{m_B^2}\ ,
\end{align}
where $m_B$ is the typical baryon mass. The baryon density just before the onset of the second inflation (denoted as 
$t=t_i$
) can be estimated as
\begin{align}
m_B\frac{n_B(t_i)}{s(t_i)} \sim m_B\frac{n_B(t_f)}{s(t_f)} \sim \frac{m_B}{\langle \sigma v \rangle M_P T_f} \sim \frac{x_f m_B^2}{M_P}~~(t_i>t_f)\ ,
\end{align}
where $n_B(t)$ and $s(t)$ denote the baryon number density and entropy density, $t_f$ and $T_f$ are the freeze-out time and temperature, respectively, and $x_f \equiv m_B / T_f$.
After the second inflation and the subsequent reheating epoch, the baryon number will be diluted to
\begin{align}
n_B(t_R) \sim n_B(t_i) \left(\frac{a(t_i)}{a(t_e)}\right)^3 \left(\frac{a(t_e)}{a(t_R)}\right)^3\ .
\end{align}
Here, 
$t_e$ and $t_r$
denote the times corresponding to the end of inflation and the completion of reheating, respectively.
The first factor accounts for dilution during inflation, 
while the second accounts for the matter-dominated era caused by inflaton oscillations between $t_e$ and $t_R$.
The scale factor ratio during inflation is related to the number of $e$-folds $N_e$ as
\begin{align}
\left(\frac{a(t_i)}{a(t_e)}\right)^3 \sim e^{-3N_e}\ .
\end{align}
During the matter-dominated reheating phase ($a \propto t^{2/3}$), the scale factor ratio can be estimated as
\begin{align}
\left(\frac{a(t_e)}{a(t_R)}\right)^3 \sim \frac{t_e^2}{t_R^2} \sim \frac{T_R^4}{T_i^4}\ ,
\end{align}
where $T_i$ is the temperature of the universe just before the onset of the second inflation and $T_R$ is the temperature at 
the end of reheating. In deriving the above equation, we have also  used approximate relations $T_i^4 / M_P^2 \sim H_I^2 \sim 1 / t_e^2$, assuming the Hubble parameter remains approximately constant during and immediately after the inflation.

\begin{figure}[!htbp]
    \centering
\includegraphics[width=1\textwidth]{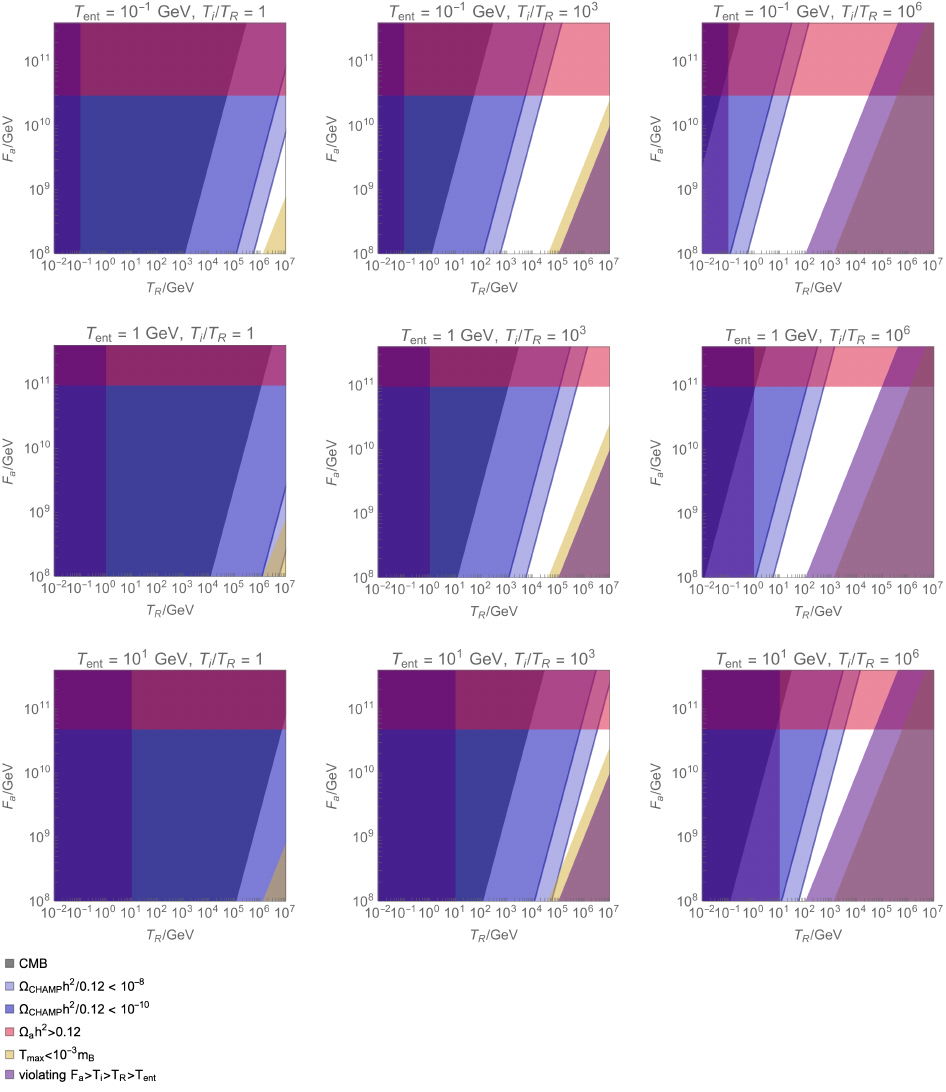}
    \caption{Viable parameter space in the $T_R$--$F_a$ plane. 
The color coding is as follows: 
\textbf{Gray}: CMB constraint on $\Omega_{\rm CHAMP} h^2$, 
\textbf{Light Blue}: $\Omega_{\rm CHAMP} h^2<10^{-8}$,
\textbf{Blue}: $\Omega_{\rm CHAMP} h^2<10^{-10}$, 
\textbf{Red}: $\Omega_a h^2 > 0.12$, 
\textbf{Brown}: $T_{\rm max} > 10^{-3} m_B$,  
\textbf{Purple}: scale ordering different from  $F_a > T_i > T_R > T_{\rm ent}$. 
The lower bound of $F_a\gtrsim 10^8\,{\rm GeV}$ is set by astrophysics~\cite{Carenza:2019pxu,Buschmann:2021juv,Springmann:2024ret,Caputo:2024oqc}.
}
\label{fig:summary}  
\end{figure}

For later convenience, we introduce the reentry temperature $T_{\rm ent}$, defined as the temperature at which cosmic strings reenter the horizon.
Assuming the strings follow the scaling law before the second inflation, there is roughly one string per Hubble patch. During inflation, these strings are diluted and their separation grows beyond the horizon. After inflation, the horizon grows faster than the comoving separation, and the strings eventually reenter.
The reentry temperature is estimated as~\cite{Harigaya:2022pjd}
\begin{align}
e^{-N_e} \sim \left( \frac{T_R^4}{T_i^4} \right)^{1/6} \frac{T_{\rm ent}}{T_R}\ .
\end{align}

Combining these relations, energy-to-entropy ratio of exotic baryons after dilution becomes
\begin{align}
\label{eq:yield}
m_B \frac{n}{s} \sim \frac{x_f m_B^2}{M_P} \left( \frac{T_{\rm ent}}{T_i} \right)^3\ .
\end{align}
In the following, we 
explore the parameter space where baryons are sufficiently diluted to satisfy multiple constraints, while ensuring that cosmic strings reenter the horizon near the QCD transition.

\subsection{Viable Parameter Space}

We now explore the parameter space in which a secondary mini-inflation effectively dilutes the relic abundance of CHAMPs, while maintaining the post-inflationary scenario for the cosmic string–wall network.
If the strings reenter the horizon  much before the QCD transition, the dynamics 
closely reproduce those of the standard post-inflationary axion scenario.  However it turns out, that efficient dilution of the exotic baryons  requires string-wall network to enter the horizon around or after the QCD transition.
 In this case, the relic abundance of axions is 
 estimated under the assumption that the entire 
 energy stored in the string–wall network is converted into axions.
The energy densities of strings and domain walls at $T = T_{\rm ent}$ (temperature when strings enter the horizon)~\cite{Harigaya:2022pjd} are given by
\begin{align}
&    \rho_{\rm string}(T_{\rm ent}) \sim   F_a^2 H^2 \log\!\left(\frac{F_a}{H}\right), \\
 &   \rho_{\rm wall}(T_{\rm ent}) \sim  F_a^2 m_a(T_{\rm ent})\,H .
 \end{align}
Here, $m_a(T_{\rm ent})$ denotes the axion mass at $T = T_{\rm ent}$ and we take it to be equal to
\bea
m_a(T)=
\min\Big[10^{-3.5} \frac{\Lambda_{\rm QCD}^2}{F_a}\Big(\frac{T}{\Lambda_{\rm QCD}}\Big)^{-4},~5.7\mu {\rm eV}\Big(\frac{10^{12}{\rm GeV}}{F_a}\Big)\Big],
\eea
where
the temperature dependence in the former expression is given from dilute instanton gas approximation, and the latter is the axion mass at the zero temperature~\cite{GrillidiCortona:2015jxo}.

Once the axion mass becomes comparable to the Hubble expansion rate,  
the axion field begins to oscillate at \( T = T_{\rm osc} \), defined by \( m_a(T) = 3H(T) \). For $T_{\rm ent}<T_{\rm osc}$, the string-wall network enters the horizon after the wall formation.
Since the wall energy dominates over the string energy soon after $T = T_{\rm ent}$, 
the string–wall system rapidly collapses at that time. 
The axion number density at the onset of collapse can then be estimated as 
$(\rho_{\rm string} + \rho_{\rm wall})/m_a(T_{\rm ent})$, 
and the present axion energy-to-entropy ratio is given by
\begin{align}
    \frac{\rho_a}{s}\Big|_{\rm present}
    = \frac{m_a(0)}{m_a(T_{\rm ent})}
    \frac{\rho_{\rm string}(T_{\rm ent}) + \rho_{\rm wall}(T_{\rm ent})}{s(T_{\rm ent})}\lesssim 0.4{\rm eV },
\end{align}
where $m_a(0)$ denotes the axion mass at the zero temperature.
On the other hand, when $T_{\rm ent}\geq T_{\rm osc}$, 
the strings can enter the horizon before domain-wall formation, 
and the dynamics are therefore close to those in the standard post-inflationary scenario.
Assuming that the dependence on $T_{\rm ent}$ is mild in this regime, 
we evaluate $\rho_a / s |_{\rm present}$ with $T_{\rm ent} = T_{\rm osc}$ for $T_{\rm ent} \geq T_{\rm osc}$.
Axions radiated from the string network may also contribute comparably. Recent numerical studies support this possibility~\cite{Gorghetto:2020qws,Buschmann:2021sdq,Kim:2024wku,Saikawa:2024bta}, while a different conclusion was reached in Ref.~\cite{Hindmarsh:2021zkt}. This contribution can lead to an $\mathcal{O}(1)$ smaller $F_a$ for the observed dark matter abundance.

To identify the viable parameter space, we consider four parameters: $F_a$, $T_{\rm ent}$, $T_R$, and $T_i$, and impose the following constraints:
\begin{align}
\label{eq:inequalities}
    &\Omega_{\rm CHAMP}h^2/0.12 < 10^{-(8-10)},~
    \Omega_a h^2/0.12 < 1, \nonumber\\
   & T_{\rm max,o} < 10^{-3} m_B, 
    ~F_a > T_i > T_R > T_{\rm ent}.
\end{align}
The first inequality comes from the current experimental constraints on the CHAMP relic density abundance. It 
comes 
from ionizing particle searches, which provide the strongest bound in the parameter space relevant for the composite axion models (i.e. $10^{8}\,\mathrm{GeV} \lesssim m_B \lesssim 10^{11}\,\mathrm{GeV}$ and  $q\sim 1$). In particular, the ICRR~\cite{Kajino:1984ug,Barish:1987zv} and BAKSAN~\cite{Alekseev:1983hpa,Alekseev:1985} experiments place limits on the CHAMP flux~\cite{Dunsky:2018mqs}.%
This translates into the maximum contribution to the relic density in the range $10^{-(8-10)}$~\cite{Dunsky:2018mqs}
\footnote{
This constraint might be relaxed if CHAMPs with sufficiently small gyro-radius cannot 
penetrate into the Galaxy due to magnetic fields as was argued in
Ref.~\cite{Chuzhoy:2008zy}, However recently these claims have been criticized in the Ref.~\cite{Perri:2025tvq}, more detailed analysis of this effect is beyond the scope of the present work.
}
and  we indicate both of these lines on the Fig.\ref{fig:summary}.
\footnote{The MACRO experiment also puts a bound on CHAMPs~\cite{MACRO:2004iiu}, although they require relatively large velocities $\beta\gtrsim 0.25$, which is not relevant in our parameter space of interest. 
The monopole search from the NOvA experiment~\cite{NOvA:2020qpg} could also be sensitive to CHAMPs, although no dedicated analysis has been reported so far.
}
We also take into account a generic constraint from the cosmic microwave background (CMB), which requires the CHAMP abundance to remain below the percent level~\cite{Dubovsky:2003yn,Burrage:2009yz,McDermott:2010pa,Dolgov:2013una,Boddy:2018wzy}. 

The second inequality  in Eq.~\ref{eq:inequalities} is a condition  that the axions do not overclose the universe. The third inequality is needed to ensure 
that
thermal production of 
baryons remains sufficiently suppressed   after the  secondary inflation.
The maximum temperature during the inflaton oscillation era of the secondary inflation can be estimated  to be
$
    T_{\rm max,o} \sim \sqrt{T_i T_R}\ ,
$
and we take
$
    m_B = 4 \pi F_a\ .
$
Then numerical factor $10^{-3}$ sufficiently suppresses the thermal production. Finally,
the hierarchy
$
    F_a > T_i > T_R > T_{\rm ent}
$
ensures a physically consistent ordering of the relevant scales.

Fig.~\ref{fig:summary} shows the viable parameter space in the $T_R$--$F_a$ plane. 
The nine panels correspond to different choices of parameters: from left to right $T_i/T_R = 1, 10^3, 10^6$, and from top to bottom $T_{\rm ent} = 0.1, 1, 10$~GeV. 
The color coding is explained in the figure caption, with the uncolored regions indicating the viable parameter space.
From these results, we identify regions where the relic abundance of heavy CHAMPs is sufficiently diluted by a secondary inflationary epoch, while standard post-inflationary axion dynamics is realized through string-wall reentry. 
This shows that a second inflation can simultaneously resolve the heavy relic problem in composite axion models and maintain the predictive features of axion cosmology. 
Further refinements--such as a more detailed treatment of string-wall dynamics and CHAMP constraints--can sharpen the boundaries of the viable region.

\section{Discussion and Conclusions}
\label{sec:outlook}
In this work, we have taken a first step toward developing a concrete framework for composite axions in the post-inflationary universe--a regime that, despite its importance, has received little attention so far. Our analysis highlights that cosmological consistency of such models hinges on two central challenges: the domain wall problem and the fate of exotic heavy relics. 
We have shown that the Lazarides–Shafi mechanism can render the domain wall number effectively 
unity in chiral gauge theories, and also a phase of mild inflation dilutes unwanted relics. Beyond reproducing the standard expectations of post-
inflationary axion cosmology, our setup predicts qualitatively new dynamics: notably, string–wall 
networks that can reenter the horizon even after the QCD transition, potentially leaving distinctive imprints.

Several open questions remain. In the cosmological context, the 
precise nature of string-wall formation and evolution cannot be 
established without numerical simulations. For instance, in models with a gauged $U(1)_{\rm 
PQ}$, both local and non-local strings can arise. Even if the domain wall number is formally 
one, non-local strings attached to multiple domain walls can reintroduce the domain wall 
problem if they remain stable below the QCD transition. Existing simulations focus on non-
composite gauged $U(1)_{\rm PQ}$ models with specific charge 
assignments and hierarchical symmetry-breaking scales, making 
it difficult to directly apply those results to our framework. 
Closely related is the determination of the axion dark matter abundance from string–wall decay, which requires dedicated simulations to establish robust predictions.
In addition, models involving non-Abelian chiral gauge dynamics 
suffer from intrinsic non-
perturbative uncertainties. In 
particular, identifying the 
spontaneous breaking patterns of 
baryon symmetries and the resulting low-energy spectrum 
requires lattice simulations, which remain technically 
challenging.
Our framework also leads to 
several potential observational 
signatures. One is the existence 
of CHAMPs, whose distribution in 
our galaxy is an outstanding 
problem and is currently being 
explored in a variety of 
experiments and observations. 
Another is the production of 
gravitational waves from a first-
order confinement phase 
transition, which could provide a 
distinctive observational window 
into our models.

\section*{Acknowledgements}
We would like to acknowledge support by the European Union - NextGenerationEU, in the
framework of the PRIN Project “Charting unexplored avenues in Dark Matter” (20224JR28W).
M.S. is supported by the MUR projects 2017L5W2PT.
M.S. thanks Yuichiro Nakai for helpful discussions,  Christiane Scherb for bringing relevant papers on CHAMP constraints to attention, Keisuke Harigaya for informing us about the CHAMP constraints. We also thank Martin Frank for discussion.

\appendix

\section{MAC understanding of the automatic axion model}
\label{app:MAC}
Let us start with a generic model based of the chiral SU(5) gauge theory with $N$ flavors of fermions $\chi_i\sim\mathbf{10},\,\psi_i\sim\mathbf{\bar{5}}$. The global symmetry of this model is $SU(N)_{\mathbf{10}}\times SU(N)_{\mathbf{\bar{5}}}\times U(1)_B$ where the fermions have 
representations $\chi\sim\mathbf{(N, 1)_1},\psi\sim\mathbf{(1,{N})_{-3}}$.
From the first 
principles we do not know how the condensation and the symmetry 
breaking will proceed, but we can use MAC (the most attractive 
channel)~\cite{Raby:1979my}.
In this case we identify the fermion condensates by looking at the interaction mediated by the tree level gauge boson exchange.
Then the largest interaction (most attractive) will correspond to the
\bea
\max[\Delta C=C_c-C_1-C_2]
\eea
where $C_{i}$ are quadratic Casimir invariants for the constituent $C_{1,2}$ and combined $C_{c}$ systems. In our case
\bea
C(\mathbf{\bar 5})=\frac{24}{5},\quad C (\mathbf{10})=\frac{36}{5}
\eea
Then for $\mathbf{\bar 5},~\mathbf{10}$ model the MAC corresponds to to
\bea
\mathbf{10}\otimes \mathbf{10}\to \mathbf{\bar 5},~~~\Delta C=\frac{48}{5}\nn
\mathbf{\bar 5}\otimes \mathbf{10} \to \mathbf{5},~~~\Delta C=\frac{36}{5}
\eea
In the first case the symmetry breaking should be similar to the one given by a scalar with quantum numbers under $SU(5)\times SU(N)_{\mathbf{10}}\times SU(N)_{\mathbf{\bar 5}}$
\bea
&&\Phi_{\mathbf{10}\times\mathbf{10}\to \mathbf{\bar 5}}:(\mathbf{\bar 5},N(N+1)/2,1)\nn
&&\Phi_{\mathbf{\bar 5}\times \mathbf{10}\to  \mathbf{5}}:(\mathbf{ 5},N, N).
\eea
The we can see that MAC  indicates the breaking of $SU(5)$ and appearance of the nonzero condensates
\bea
\langle \mathbf{10} \mathbf{10}\rangle,~~\langle \mathbf{\bar 5}\mathbf{10}\rangle \neq 0.
\eea

\section{More on chiral moose composite axion}
\label{app:more_on}

\begin{table}[h!]
    \centering
    \renewcommand{\arraystretch}{1.5}
    \begin{tabular}{|c|c|c|c|c|c|c|c|c|}
        \hline
        Field & $[SU(3)_0]$ & $SU(4)_L^{\rm approx}$ & $[SU(5)_1]$ & \textbf{$[SU(4)_1]$} &  $[SU(5)_2]$ &  $SU(4)_R^{\rm approx}$ & $U(1)_{\rm PQ}$ & $U(1)_{\rm PQ'}$ \\ \hline
        $Q^\alpha_A$ &  $\mathbf{3}$ & \multirow{2}{*}{$\mathbf{4}$} & $\bar{\mathbf{5}}$ &  & & & -3 & 1   \\ 
        \cline{1-2} \cline{4-9}
        $\eta_A$ & $\mathbf{1}$ & & $\bar{\mathbf{5}}$ &  &  & & -3  & -3\\ \hline
        $\chi_{1,1}^{AB\,\hat\alpha}$ &  & & $\mathbf{10}$ & $\mathbf{4}$ &  & & 1 & 0  \\ \hline
        $\psi_{1,2\,\hat\alpha}^{\hat A\hat B}$ &  & & & $\mathbf{\bar 4}$ &  $\mathbf{10}$ & & 0  & 0\\ \hline
         $\bar Q_{\alpha\hat A}$ &  $\mathbf{\bar 3}$  &  & &  & $\mathbf{\bar 5}$ &  \multirow{2}{*}{$\bar{\mathbf{4}}$} & 0 & 0   \\ \cline{1-6}\cline{8-9}
        $\bar\eta_{\hat A}$ & $\mathbf{1}$ &  &  & & $\bar{\mathbf{5}}$ & & 0  & 0\\ \hline
    \end{tabular}
    \caption{Matter contents.}
    \label{tab:matter_contents_app}
\end{table}

In the chiral moose model, 
there appears to be an additional Peccei-Quinn symmetry \( U(1)_{\rm PQ'} \), as shown in Table~\ref{tab:matter_contents_app}, which is anomalous under \( [SU(3)_0]^2 \) but anomaly-free with respect to \( [SU(5)_1]^2 \). 
However, one linear combination of $U(1)_{\rm PQ}$ and $U_{\rm PQ'}$ can be made anomaly-free under both \( [SU(3)_0] \) and \( [SU(5)_1] \), and hence there exists one PQ symmetry up to anomaly-free redefinitions.

More concretely, 
the anomaly coefficients of each $U(1)_{\rm PQ}$ and $U_{\rm PQ'}$ with $SU(3)_c$ are given by
\begin{align}
  \mathcal{A}(U(1)_{\rm PQ}-SU(3)_c^2) &= -3 \times 5 + 1 \times 10 = -5, \\
  \mathcal{A}(U(1)_{\rm PQ'}-SU(3)_c^2) &= 1 \times 5 = 5.
\end{align}
We may define a new  \( U(1)_{\text{free}} \) symmetry with charges given by the sum of the charges under \( U(1)_{\rm PQ} \) and \( U(1)_{\rm PQ'} \), i.e. , \( q_{U(1)_{\rm PQ}} + q_{U(1)_{\rm PQ'}} \). This new combination is anomaly-free under the gauge groups:
\[
 \mathcal{A}(U(1)_{\text{free}} - [SU(3)_0]^2) = 0, \quad \mathcal{A}(U(1)_{\text{free}} - [SU(5)_1]^2) = 0.
\]
This implies that \( U(1)_{\rm PQ'} \) can be written as a linear combination of \( U(1)_{\rm PQ} \) and \( U(1)_{\text{free}} \). 

Let us now consider the following condensates discussed in the main text:
\begin{align}
    \langle \mathcal{\chi}_{1,1} \mathcal{\chi}_{1,1} \mathcal{\chi}_{1,1} \Psi_1 \rangle &\neq 0, \\
    \langle \mathcal{\chi}_{1,1} \Psi_1 \Psi_1 \, \mathcal{\psi}_{1,2} \Psi_2 \Psi_2 \rangle &\neq 0,
\end{align}
where \( \Psi_1 \) includes both \( Q \) and \( \eta \), while \( \Psi_2 \) includes both \( \bar{Q} \) and \( \bar{\eta} \). Gauge and spinor indices are contracted appropriately. The first operator is charged under \( U(1)_{\text{free}} \) but neutral under \( U(1)_{\rm PQ} \), while the second operator carries charges under both \( U(1)_{\text{free}} \) and \( U(1)_{\rm PQ} \). Once the first condensate forms and spontaneously breaks \( U(1)_{\text{free}} \), there is no remaining freedom to perform a \( U(1)_{\text{free}} \) rotation. The condensation of the second operator then breaks \( U(1)_{\rm PQ} \). These observations suggest that it is more convenient to identify \( U(1)_{\rm PQ} \) as the PQ symmetry, rather than \( U(1)_{\rm PQ'} \), as mentioned in the main text.

\subsection*{Extending the moose diagram}

\begin{table}[!htbp]
    \centering
    \renewcommand{\arraystretch}{1.5}
    \begin{tabular}{|c|c|c|c|c|c|c|c|}
        \hline
         $SU(3)_0$ & $SU(5)_1$ & \textbf{$SU(4)_1$} &  $SU(5)_2$ & $SU(4)_2$ & $SU(5)_3$  & $SU(4)_3$  & $SU(5)_4$ \\ \hline
          $\mathbf{3}$ & $\bar{\mathbf{5}}$ &  & & & & &\\ \hline
        $\mathbf{1}$ & $\bar{\mathbf{5}}$ &  &  &  &  & &\\ \hline
          &  $\mathbf{10}$ & $\mathbf{4}$ &  &  &  & &\\ \hline
         &  & $\mathbf{\bar 4}$ &  $\mathbf{10}$ &  &  & & \\ \hline
           &  &  & $\mathbf{\bar 5}$  & $\mathbf{\bar 4}$ &  & &  \\ \hline
        &  &  &  & $\mathbf{4}$ & $\mathbf{\bar 5}$   & &\\ \hline
            &  &  &  &  & $\mathbf{10}$  &  $\mathbf{4}$&  \\ \hline
         &  &  &  &  &   & $\mathbf{\bar 4}$ & $\mathbf{10}$\\ \hline
       $\mathbf{\bar 3}$  &  &  &  &  &   &  & $\mathbf{\bar 5}$\\\hline
        $\mathbf{1}$  &  &  &  &  &   &  & $\mathbf{\bar 5}$\\\hline
    \end{tabular}
    \caption{Matter contents. $\alpha$ denotes $SU(3)_0$ indices. $A,B$ denotes the $SU(5)_1$ indices. $\hat A,\hat B$ denote the $SU(5)_2$ indices.}
    \label{tab:matter_contents_app_deconstruction}
\end{table}

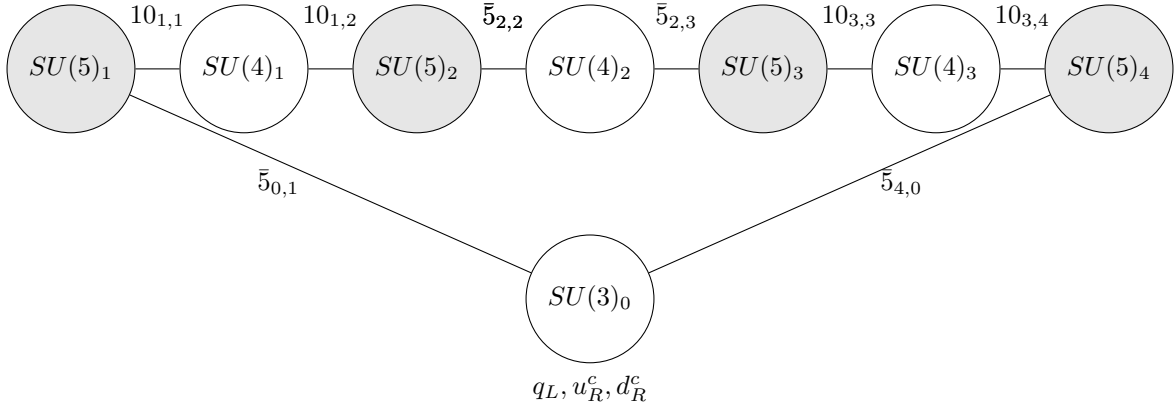
\begin{figure}[!htbp] 
  \centering
\begin{tikzpicture}[xscale=2.3,yscale=1.7, 
every node/.style={font=\footnotesize}
]
  \node[circle, draw, fill=gray!20,  minimum size=1.7cm, inner sep=0pt] (N1) at (0,0) {$SU(5)_1$};
  \node[circle, draw,  minimum size=1.7cm, inner sep=0pt] (N2) at (1,0) {$SU(4)_1$};
  \node[circle, draw,  fill=gray!20, minimum size=1.7cm, inner sep=0pt] (N3) at (2,0) {$SU(5)_2$};
  \node[circle, draw,  minimum size=1.7cm, inner sep=0pt] (N4) at (3,0) {$SU(4)_2$};
   \node[circle, draw,  fill=gray!20, minimum size=1.7cm, inner sep=0pt] (N5) at (4,0) {$SU(5)_3$};
    \node[circle, draw,  minimum size=1.7cm, inner sep=0pt] (N6) at (5,0) {$SU(4)_3$};
   \node[circle, draw,  fill=gray!20, minimum size=1.7cm, inner sep=0pt] (N7) at (6,0) {$SU(5)_4$};
  
  \node[circle, draw, minimum size=1.7cm, inner sep=0pt] (SU3) at (3,-1.8) {$SU(3)_0$};
  
  \node at (0.5,0.4) {$10_{1,1}$};
  \node at (1.5,0.4) {$10_{1,2}$};
  \node at (2.5,0.4) {$\bar 5_{2,2}$};
  \node at (2.5,0.4) {$\bar 5_{2,2}$};
  \node at (3.5,0.4) {$\bar 5_{2,3}$};
  \node at (4.5,0.4) {$10_{3,3}$};
  \node at (5.5,0.4) {$10_{3,4}$};

  \draw[-] (N1) -- (N2);
  \draw[-] (N2) -- (N3);
  \draw[-] (N3) -- (N4);
  \draw[-] (N4) -- (N5);
  \draw[-] (N5) -- (N6);
  \draw[-] (N6) -- (N7);
  
  \draw[-] (N1) -- (SU3);
  \draw[-] (N7) -- (SU3);

  \node at (1.2,-0.9) {$\bar 5_{0,1}$};
  \node at (4.8,-0.9) {$\bar 5_{4,0}$};
  \node at (3,-2.5) {$q_L, u^c_R, d^c_R$};
  
\end{tikzpicture}
\caption{Moose diagram for extending the chiral moose model. The link fermion fields are denoted by their $SU(5)$ representations.}
  \label{fig:moose_deconstruct_app}
\end{figure}

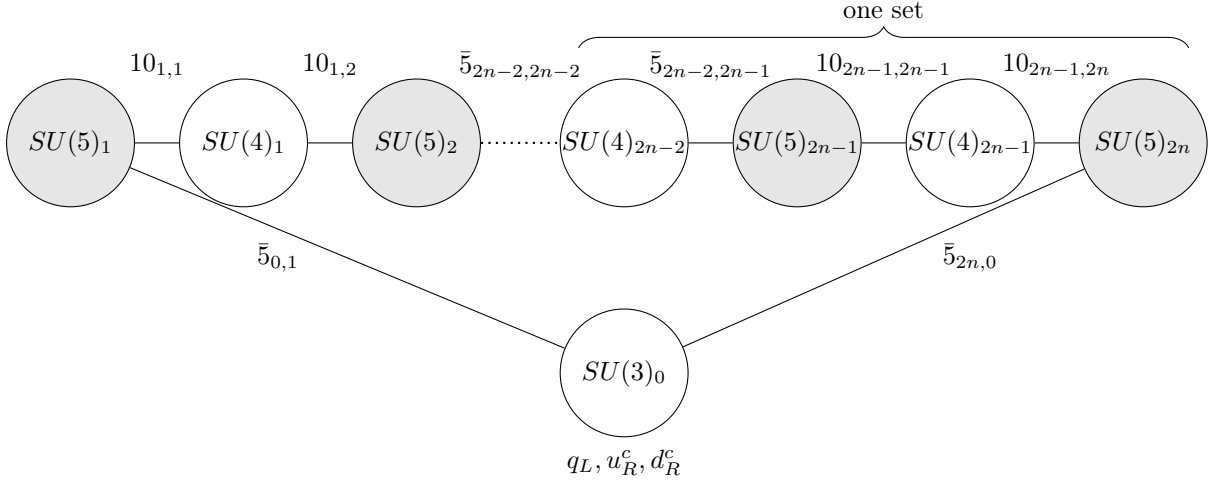
\begin{figure}[!htbp] 
\begin{tikzpicture}[xscale=2.3,yscale=1.7, 
every node/.style={font=\footnotesize}
]
  \node[circle, draw, fill=gray!20,  minimum size=1.7cm, inner sep=0pt] (N1) at (0,0) {$SU(5)_1$};
  \node[circle, draw,  minimum size=1.7cm, inner sep=0pt] (N2) at (1,0) {$SU(4)_1$};
  \node[circle, draw,  fill=gray!20, minimum size=1.7cm, inner sep=0pt] (N3) at (2,0) {$SU(5)_2$};
  \node[circle, draw,  minimum size=1.7cm, inner sep=0pt] (N4) at (3.2,0) {$SU(4)_{2n-2}$};
   \node[circle, draw,  fill=gray!20, minimum size=1.7cm, inner sep=0pt] (N5) at (4.2,0) {$SU(5)_{2n-1}$};
    \node[circle, draw,  minimum size=1.7cm, inner sep=0pt] (N6) at (5.2,0) {$SU(4)_{2n-1}$};
   \node[circle, draw,  fill=gray!20, minimum size=1.7cm, inner sep=0pt] (N7) at (6.2,0) {$SU(5)_{2n}$};
  
  \node[circle, draw, minimum size=1.7cm, inner sep=0pt] (SU3) at (3.2,-1.8) {$SU(3)_0$};
  
  \node at (0.5,0.6) {$10_{1,1}$};
  \node at (1.5,0.6) {$10_{1,2}$};
  \node at (2.6,0.6) {$\bar 5_{2n-2,2n-2}$};
  \node at (3.7,0.6) {$\bar 5_{2n-2,2n-1}$};
  \node at (4.7,0.6) {$ 10_{2n-1,2n-1}$};
  \node at (5.7,0.6) {$ 10_{2n-1,2n}$};

  \draw[-] (N1) -- (N2);
  \draw[-] (N2) -- (N3);
  \draw[dotted, thick] (N3) -- (N4);
  \draw[-] (N4) -- (N5);
  \draw[-] (N5) -- (N6);
  \draw[-] (N6) -- (N7);
  \draw[-] (N1) -- (SU3);
  \draw[-] (N7) -- (SU3);
  \draw[decorate,decoration={brace,amplitude=8pt}]
  ([yshift=11pt] N4.north west) -- ([yshift=11pt] N7.north east)
  node[midway,yshift=+0.5cm]{one set};

  \node at (1.2,-0.9) {$\bar 5_{0,1}$};
  \node at (5.2,-0.9) {$\bar 5_{2n,0}$};
  \node at (3.2,-2.5) {$q_L, u^c_R, d^c_R$};
  
\end{tikzpicture}
\caption{Moose diagram for further extension of the chiral moose model $(n>1)$.}
  \label{fig:moose_deconstruct_app_2}
\end{figure}

Our model can be extended in a manner similar to a vector-like moose model.
As a first step, we illustrate a simple extension by adding $SU(4)_{2,3}$ and $SU(5)_{3,4}$ gauge sectors as summarized in Table~\ref{tab:matter_contents_app_deconstruction} and shown in the moose diagram of Fig.~\ref{fig:moose_deconstruct_app}. 
Further extensions lead to a chain-like structure, which is represented by a moose diagram, as shown in Fig.~\ref{fig:moose_deconstruct_app_2}.
Here, we successively add sets consisting of two $SU(4)$ and two $SU(5)$ sectors (indicated by the bracket in Fig.~\ref{fig:moose_deconstruct_app_2}).
This deconstruction-inspired framework provides a more flexible platform for addressing the axion quality problem.
In particular, by distributing the PQ-breaking dynamics across multiple sites, one can more effectively suppress dangerous higher-dimensional operators.
A detailed investigation of the deconstructed setup and its relation to extra-dimensional realizations is left for future work.

\section{Running of QCD gauge coupling}
\label{app:beta}
One of the key successes of QCD is its asymptotic freedom and the strong regime of its gauge coupling at low energy which agrees with hadron physics \cite{Gross:1973, Politzer:1973}. This is represented through its beta function (to one-loop order)
\bea
\beta(g_s) = -\frac{g^3_s}{16 \pi^2}\left(\frac{11}{3}\,N_c-\frac{2}{3}N_f\right)
\label{qcd_beta}
\eea
where $N_f$ denotes the number of Dirac fermion flavors in the fundamental representation of $SU(3)_c$. The asymptotic freedom $\beta(g_s)<0$ requires that the number of flavors satisfy $N_f\leq16$. The standard model has $6$ flavors of quarks. The models of composite axion necessarily include additional colored fermions that should not exceed $10$ flavors. \\ \\
We consider the models discussed in the paper. First, the minimal Kim-Choi model \ref{sec:kim} adds $N$ "flavors" of fermions $Q$ in the representation $\mathbf{3}$ of $SU(3)_c$. Then it is required that $N\leq 10$. The automatic PQ model \ref{sec:automatic} built on gauge group $SU(5)$ adds $5$ "flavors" of fermions $\bar{F}$ and $10$ "flavors" of fermions $A$. This leads to clear violation of the asymptotic freedom requirement. \\ \\
The gauged PQ-model \ref{sec:gauged_pq} adds $N+M$ flavors from both sectors. The asymptotic freedom constraint requires $N+M\leq 10$, while the model requires ${\rm gcd} (N,M)=1$.

\bibliographystyle{JHEP}
\bibliography{ref}

\end{document}